\documentclass[preprint]{aastex631}
\shorttitle{diffusion coefficients around supernova remnants}
\shortauthors{Bao, Blasi \& Chen}
\graphicspath{{./}{figures/}}

\begin{document}

\title{Regions of suppressed diffusion around supernova remnants?}

\author[0000-0001-8918-5248]{Yiwei Bao}
\affil{Department of Astronomy, Nanjing University, 163 Xianlin Avenue, Nanjing 210023, China}
\affiliation{Gran Sasso Science Institute, Viale F.\ Crispi 7, 67100 L’Aquila, Italy}

\author[0000-0003-2480-599X]{Pasquale Blasi}
\affiliation{Gran Sasso Science Institute, Viale F.\ Crispi 7, 67100 L’Aquila, Italy}
\affiliation{INFN/Laboratori Nazionali del Gran Sasso, Via G. Acitelli 22, Assergi (AQ), Italy}

\author[0000-0002-4753-2798]{Yang Chen}
\affil{Department of Astronomy, Nanjing University, 163 Xianlin Avenue, Nanjing 210023, China}
\affiliation{Key Laboratory of Modern Astronomy and Astrophysics, Nanjing University, Ministry of Education, Nanjing, China}

\correspondingauthor{Yiwei Bao}
\email{yiwei.bao@gssi.it}
\correspondingauthor{Pasquale Blasi}
\email{pasquale.blasi@gssi.it}
\correspondingauthor{Yang Chen}
\email{ygchen@nju.edu.cn}

\begin{abstract}
The recent discovery of the so-called TeV halos has attracted much attention. The morphology of the emission requires that the region is characterized by severe suppression of the diffusion coefficient. This finding raises many questions as to its origin: 1) is the suppressed diffusion { to be} attributed to instabilities induced by the same radiating particles? 2) or does it actually show that the diffusion coefficient is small throughout the disc of the Galaxy? In both cases, one would expect that the surroundings of supernova remnants (SNRs) should also show evidence of reduced diffusion coefficient, since most remnants are located in the disc and are expected to be sites of effective particle acceleration. Should we expect the existence of regions of extended $\gamma$-ray emission from these regions as well? { Here we investigate the transport of cosmic rays (CRs) escaped from SNRs in order to assess the viability of the idea of having a cocoon of suppressed diffusion around them. A comparison of our results with the $\gamma$-ray emission from the regions around HB9 and W28 does not provide solid evidence of reduced diffusivity. However, if indeed the phenomenon of reduced diffusivity occurs around SNRs surrounded by molecular clouds, our calculations show that the effects on the grammage of Galactic CRs can be significant.} 
\end{abstract}


\section{Introduction} \label{sec:Introduction}

The recent discovery of the so-called TeV halos \citep[see e.g.,][]{2017Sci...358..911A,2019PhRvD.100d3016S}, regions of spatially extended $\gamma$-ray emission at energies $\gtrsim$ TeV around selected pulsar wind nebulae (PWNe), has raised a series of questions that might have a deep impact on our conventional picture of the transport of cosmic rays (CRs). The morphology of these halos is such that the emission is best interpreted as inverse Compton scattering (ICS) of the electrons and positrons originated from the PWN, propagating diffusively in a region of size $\sim 50$ pc. The diffusion coefficient in the region appears to be suppressed by a factor $\sim 100-1000$ with respect to the value deduced for the ISM using the standard indicators, such as the secondary to primary ratio \citep[see e.g.,][]{2016PhRvL.117w1102A} and the abundance of unstable isotopes \citep[see e.g.,][]{2007ARNPS..57..285S}. A detailed description of the transport of electron-positron pairs in the surroundings of the Geminga pulsar confirms the need for a reduced diffusion coefficient \cite[]{SchroerGeminga2023}. 

This discovery raised a series of questions with potentially crucial implications for the standard model of the origin of Galactic CRs: 1) Is the diffusion coefficient small only around PWNe or everywhere in the Galactic disc? 2) A small diffusion coefficient in the $\gtrsim$ TeV energy range implies the existence of a large level of turbulence, or a small coherence length of the turbulence (or both), but is this turbulence extrinsic or rather caused by CRs through the excitation of streaming instabilities? 3) What are the implications of the suppressed diffusion coefficient around sources for the grammage (the mass traversed by CRs during their propagation) and the positron excess? 

These questions are not all independent from each other: for instance, the answer to the third question depends very sensibly upon whether the diffusion coefficient is only small around PWNe or also around SNRs, and it also depends on whether the diffusion coefficient is small everywhere in the disc or only in cocoons of limited size around sources. Moreover, the possibility for CRs to excite streaming instability and lead to reduced diffusivity depends on whether the particles are protons from a SNR or rather a quasi-neutral electron-positron beam from PWNe, or even a spatially charge-separated beam of very high energy electrons or positrons from PWNe \cite[]{2019MNRAS.488.5690O,Olmi2024}. 

Here we focus on some of these questions: we investigate the information that can be gathered from $\gamma$-ray observations of the regions surrounding selected SNRs, in the attempt to infer whether there is evidence of suppressed diffusivity there too, as would be expected if the diffusion coefficient is small everywhere in the Galactic disc. Moreover, since the self-generation of turbulence induced by CR streaming is related to the luminosity of the source, studying SNRs with different intrinsic energy content should lead to different levels of turbulence in the surrounding medium, which should in turn reflect in different $\gamma$-ray emission. 

In order to start addressing these issues, we consider two approaches to CR diffusion. The first is a { simple phenomenological one already used by several authors before (for instance \cite{2008A&A...481..401A,2013MNRAS.429.1643N,Bao2019,Bao2021})}, in which diffusion is assumed to occur in one or three spatial dimensions, at a level that can be calibrated to fit the data. In other words the normalisation and energy dependence of the diffusion coefficient are parameters of the problem. This approach is appropriate to describe transport in a disc where the coherence length is small or there is an enhanced level of turbulence, so that spatial diffusion is slow. The approach may also be used to check the possibility that some set of data may hint at small diffusion coefficient, independent of the origin of the suppressed diffusion. 

The second approach is one in which the diffusion coefficient is an output of the calculation and the level of turbulence is due to the excitation of a CR streaming instability. Since the gradient in the number density of non-thermal particles is the largest near the sources, the propagation of CRs in these regions can be strongly affected by their self-generated turbulence, as previously pointed out by several authors \citep[see e.g.,][]{ 2008AdSpR..42..486P,2016PhRvD..94h3003D,2016MNRAS.461.3552N,2018MNRAS.474.1944D,2019MNRAS.484.2684N,2022A&A...660A..57R}. 

In the latter case, the assumption is that the pre-existing magnetic field is sufficiently ordered around the source (namely its coherence length is sufficiently large) so that transport can be, in first approximation, considered as one-dimensional. If the field around the source were disordered on scales smaller than the size of the source to start with, the problem would be 3D and the gradients would become too small to allow for self-generation, immediately outside the source region. Moreover, as discussed in previous literature, the process of wave generation induced by particle streaming becomes inefficient (even in one dimension) for particles with energies $\gtrsim$ TeV. 

It is clear that a reduced diffusion coefficient around sources (or in the disc) leads to potentially important effects in terms of grammage accumulated by CRs in the Galaxy { \cite[]{2016PhRvD..94h3003D,2022A&A...660A..57R}}. 


{ 
In the present work we focus on several new aspects: 1) we first make an attempt at framing the discussion of the phenomenon of reduced diffusivity in the context of current open issues related to the transport of CRs in the Galaxy (for instance the positron excess and alternative models of CR transport in which grammage is accumulated near sources) and in the light of the discovery of the so-called TeV halos (see discussion in Sec. \ref{sec:difference}). 2) We discuss the requirements to be imposed on clouds in the regions around SNRs in order to identify the effects of reduced diffusivity. This is an all but trivial process, as detailed in Sec. \ref{sec:observations}, and only few known SNRs remain viable after imposing these constraints. Here we focus on the two SNRs HB9 and W28 and the MCs around them. 3) We assess the role of molecular clouds (MCs) around SNRs in terms of the total grammage that CRs can accumulate around sources, a topic that has recently attracted much attention in the aftermath of alternative models of CR transport \cite[]{Cowsik2014,Lipari2017}. We carry out these calculations including the effects of different types of damping. We show that the presence of clouds and the onset of self-confinement may lead to the accumulation of grammage near sources that significantly exceeds naive expectations. 4) Finally we comment on different models of particle escape from SNRs in terms of the interpretation of the gamma ray emission from near-source regions. 

As mentioned above,} we apply the two approaches described above to two SNRs{ , HB9 and W28,} for which there is evidence of extended $\gamma$-ray emission from MCs situated in the region around them and illuminated by CRs accelerated at the SNRs, a situation that has been suggested to be ideal for the investigation of CR escape from sources \citep[see e.g.,][]{2008A&A...481..401A}. 
{ The first SNR we focus on, HB9,} is known to be an under-luminous source and is chosen because one should expect weak CR induced instabilities to be excited around such an object and nevertheless $\gamma$-ray emission has been observed from the region around it. If we can confirm that no suppression of the diffusion coefficient is required to explain such $\gamma$-ray emission, then it would suggest that the small diffusivity is not a property of the Galactic disc at large. The second SNR, W28, has long been claimed to sit in a region with possibly reduced diffusion coefficient \citep[see e.g.,][] {2007Ap&SS.309..365G,2009MNRAS.396.1629G,2011MNRAS.410.1577O,2012MNRAS.421..935L}, but as we discuss below, the claim is more or less serious depending on whether diffusion is treated in one or three spatial dimensions \citep[see also][]{2009MNRAS.396.1629G}.

The article is organised as follows: in \S\ref{sec:difference}, we discuss the difference between SNRs and PWNe as sources of high energy particles { and we discuss the value of calculations such as the ones illustrated here to determine the impact of reduced diffusivity on CR transport}. In \S\ref{sec:observations}, we list some observational facts about HB9 and W28 { and discuss the reasons for selecting specific remnants and specific clouds around them to investigate the role of non-linear CR transport}. In \S\ref{sec:diffusion}, we describe our theoretical approach to the phenomenological model and the one based on self-generation. The results in terms of gamma ray emission from { specific clouds in} the regions surrounding HB9 and W28 are discussed in  \S\ref{sec:Results}. The important implications of suppressed diffusion in a region with MCs on the grammage of Galactic CRs are discussed in \S\ref{sec:Grammage}. Our conclusions are drawn in \S\ref{sec:Conclusions}.

\section{Differences and Analogies between PWNe and SNRs as Particle Accelerators}
\label{sec:difference}

{ In order to understand the role of self-generation around sources and whether this phenomenon can be responsible for  reduced diffusivity, it is important to identify analogies and differences between the two main classes of sources where these considerations are usually applied to.}

Multi-frequency observations of SNRs and PWNe leave no doubt that both classes of sources act as effective particle accelerators. However the way particles are energized, the types of particles that get accelerated and the ways the particles escape from the sources are very different in the two cases, as we discuss below. 

In SNRs, particle acceleration mainly occurs at the forward shock, where the ISM plasma is slowed down and heated up \citep[see][for reviews]{2013A&ARv..21...70B,BlasiNCim}. In an ideal situation, the accelerated particles are all advected toward downstream, both during the ejecta-dominated and the Sedov-Taylor { (ST)} phases, with the possible exception of the particles at the maximum energy. The time dependence of the maximum energy depends on many factors, such as the strength of the amplified magnetic field, as due to CR streaming instability excited upstream of the shock \citep{2004MNRAS.353..550B,2008MNRAS.385.1946A}. The integration over time of the escape flux eventually results in an approximately power-law, typically hard, spectrum \citep[see e.g.,][]{2010APh....33..160C,2014MNRAS.437.2802S,2020APh...12302492C}. The rest of the accelerated particles are liberated into the ISM after the shock has slowed down and the flow has become subsonic. In fact, it is believed that escape should take place at the beginning of the radiative phase \citep[see e.g.,][]{2020APh...12302492C}, when instabilities are hard to sustain and particles may be free to leave. However, it is fair to say that a proof of this latter point is missing and there is no observational fact in its support, with the exception of the lack of non-thermal activity from objects that are in the radiative phase. By the time of escape, particles trapped in the downstream have been confined there for tens of thousands of years, and have suffered substantial adiabatic and, for leptons, radiative energy losses \citep[]{2021A&A...650A..62C}. 

In a less ideal situation, for instance if the shock is not perfectly spherical and/or is broken at some locations, there may be leakage of the whole spectrum (namely at all energies)  from the downstream region at any time. This ambiguity is part of the difficulty in modelling the escape of accelerated particles from a SNR \citep[see discussion in][]{2021A&A...650A..62C}. 

As discussed by \citet{2020APh...12302492C}, in this basic picture, the high energy particles liberated by a SNR are mainly protons (and nuclei), while the electron spectrum is typically limited to energies $\lesssim 10$ TeV, because of energy losses inside the acceleration region and during transport downstream of the shock. This is important in terms of formation of a TeV halo around SNRs, as we discuss below. 

Regarding PWNe, particle acceleration is expected to take place mainly at the termination shock, where the pulsar wind is decelerated and heated. It is not yet known whether the process of particle acceleration is solely diffusive shock acceleration (DSA), for which the termination shock is ultra-relativistic and is usually not expected to behave as a good accelerator \cite[]{2018ApJ...863...18G}{ ,} or magnetic reconnection \citep{2021ApJ...908..147L} or a combination of the two. The spectrum of the accelerated particles, which, as inferred from multi-frequency observations, typically has the shape of a broken power-law with a change of slope at lorentz factor of the particles $\sim 10^5$, suggests that the process is not as simple as DSA alone.

It is important to realize that the observed TeV halos surround pulsars that, due to their birth kick motion, have already left their parent SNR. This is of crucial importance in that the accelerated particles may now leave the bow shock nebula that is formed due to the interaction of the pulsar wind with the ISM, through the comet-like tail that is typical of bow shock nebulae. The particles that are accelerated at the termination shock are expected to be electron-positron pairs, with an energy that can in principle be as high as the potential drop of the pulsar. This fact alone, the dominance of leptons and their escape from the acceleration region, leads to the conclusion that PWNe are more effective in injecting leptons in the surrounding region than SNRs, which in turn may explain why so far we have observed TeV halos around PWNe. 

The observed TeV halos show a roughly spherical shape. The size and morphology of the $\gamma$-ray emission are best explained if the diffusion coefficient in the region around PWNe is small, typically $100-1000$ times smaller than inferred in the whole CR propagation volume \citep[see e.g.,][]{2017Sci...358..911A,SchroerGeminga2023}. 

Three options can be considered: 1) the diffusion coefficient is small everywhere through the disc of the Galaxy, perhaps due to the larger amount of extrinsic turbulence there; 2) the diffusion coefficient is small around sources, because of CR induced streaming instability in the regions where currents are stronger; 3) the diffusion coefficient is small in some locations, for some reason, and the PWN produces a TeV halo if it happens to cross one such regions. Case 3) would also apply to SNRs with the caveat that electrons are more efficient radiators, and as stressed above, PWNe are better sources of high energy leptons than SNRs. The occurrence rate of extended TeV emission around sources, as in case 3), would reflect the probability of finding such sources in the areas of low diffusivity, and at present cannot be estimated in any reliable way.

In case 1), since both PWNe and SNRs mainly reside in the disc, there should be evidence of $\gamma$-ray emission around both classes of sources, although PWNe would shine due to leptonic ICS emission while SNRs require target gas (for instance in the form of MCs) to produce $\gamma$-ray emission. 

In case 2), the situation requires a deeper analysis: CRs escaping SNRs are known to be able to produce waves through the excitation of streaming instability \citep[see e.g.,][]{2008AdSpR..42..486P}, although the phenomenon is limited by different mechanisms of damping, depending on the environment \cite[]{2016PhRvD..94h3003D,2016MNRAS.461.3552N,2018MNRAS.474.1944D,2019MNRAS.484.2684N,2022A&A...660A..57R}. The current of CRs escaping a young SNR may even be sufficient to excite a non resonant streaming instability, as discussed by \citet{2021ApJ...914L..13S,2022MNRAS.512..233S}. 

On the other hand, the excitation of streaming instability by the pairs escaping a PWN did not receive equal attention: if the pairs are liberated into the ISM without charge separation, the total current that they carry vanishes and the non resonant branch \citep{2004MNRAS.353..550B} of the instability cannot be excited. The resonant streaming instability is excited even in a situation of zero net current \citep{2018PhRvD..98f3017E} but the resulting level of self-generated turbulence is insufficient to explain the strong suppression of diffusion coefficient, unless the transverse cross section of the flux tube is assumed to be very small \citep{2022PhRvD.105l3008M}, which is clearly at odds with the quasi-spherical morphology of the emission. Some evidence that electrons and positrons at energies comparable with the potential drop of the pulsar may be injected at different locations, thereby making the net current locally non-zero, was presented by \citep{2019MNRAS.488.5690O}. The effect of the resulting current for the excitation of the non-resonant streaming instability is currently being investigated { \cite{Olmi2024}}. This phenomenon might be related to another non-thermal unexpected occurrence, namely the long and narrow bright X-ray filaments seen around some PWNe (see for instance \cite{HuiBecker2007}). This latter phenomenon might be somehow related to the production of TeV halos at later stages.  

In the following, we will develop diffusion models in cases 1) and 2) and apply them to SNRs as exemplified by HB9 and W28. In these cases, there is evidence of extended $\gamma$-ray emission which, when projected, seems to lie outside (but near) the remnants, and the $\gamma$-ray emission appears to be coincident with large MCs, not in direct contact with the remnants' shocks.

\section{Some observational facts about HB9 and W28}
\label{sec:observations}

{ Here we provide some observational facts about SNRs HB9 and W28}. Gamma-ray emission appears to arise from several MCs around these SNRs, { but only some of this emission can plausibly be used to infer information about the transport of escaping particles in the near-source regions.}

There are two MCs in the region around HB9, one (called ``R1'') that appears to be in touch with the SNR shock { and} the other (called ``R2'') { that} is located at an angular distance from the explosion center of $\sim 1.4^\circ$ (corresponding to $\sim 15$ pc) \citep{2022PASJ...74..625O}. 

Around W28, there are four MCs, namely, MCs N (which is interacting with the shock directly) A, B, and C \citep[see e.g.,][]{2008A&A...481..401A, 2018ApJ...860...69C}. The $\gamma$-ray emission of HESS J1800$-$24\,C (corresponding to MC C) is suspected to be contaminated by the contribution from the SNR candidate G5.71$-$0.08 \citep{2006ApJ...639L..25B,2008A&A...481..401A}. Hence here we exclude from our analysis the contributions from both MCs N and C.

The $\gamma$-ray emission from the surroundings of both SNRs share some similarities in terms of spectrum. A recent Fermi-LAT data analysis of HB9 \cite[]{2022PASJ...74..625O} found that the $\gamma$-ray spectra from MCs R1 and R2 are somewhat harder (with power-law indices $\Gamma =1.84 \pm 0.18$ and $\Gamma =1.84\pm 0.14$) than that at the SNR shell ($\Gamma= 2.55 \pm 0.10$). Likewise, for W28, MCs A and B show much harder $\gamma$-ray spectra than MC N (which is interacting with the shock directly). The proton spectral indices required to fit the $\gamma$-ray spectra of MCs N, A, and B are 2.7, 2.2, and 2.4, respectively \citep[see Fig.\ 6 in][]{2018ApJ...860...69C}. Both instances can be considered as indications of a possible energy-dependent diffusion of particles to the locations of the MCs.

\subsection{HB9}

HB9 is a thermal composite (or mixed-morphology) SNR, with an estimated age $t_{\rm age }\approx 7$ kyr \citep{1995A&A...293..853L, 2009JAHH...12...61I}. A radio continuum shell at 408 MHz is found to have a good spatial correspondence with HI and CO clouds \citep{2019MNRAS.489.4300S}. The distance to the SNR is measured to be $d\approx 600$ pc \citep{2019MNRAS.489.4300S}, which is roughly consistent with the estimate (800 pc) in \citet{2007A&A...461.1013L}. The SNR has a large angular radius of $1^\circ$ ($\sim 10$ pc at 600 pc), with its shock suggested to be interacting directly with the nearby MC R1 as flagged by the detection of 1720 MHz OH maser \citep{1996AJ....111.1651F}. The density of the X-ray-emitting gas is estimated to be $\sim 1.4 f_{\rm fill}^{-1/2}(d/0.6$ kpc$)^{-1/2}$\,cm$^{-3}$ (where $f_{\rm fill}$ is the filling factor) in the eastern region and $\sim 0.9 f_{\rm fill}^{-1/2}(d/0.6$ kpc$)^{-1/2}$\,cm$^{-3}$  in the western region \citep{2019MNRAS.489.4300S}. Based on cloud evaporation models \citep{1991ApJ...373..543W}, \citet{2007A&A...461.1013L} suggested that the preshock intercloud density is $\sim (3$--$10)\times 10^{-3}$ cm$^{-3}$, with the ratio between the cloud mass $m_{\rm MC}$ and the intercloud medium mass being $\sim$ 20--50. Therefore the mean ISM density is $\sim 0.2$--$0.5$ cm$^{-3}$, which is consistent with the estimate in \citet{2019MNRAS.489.4300S}. In order to fit the radius of the SNR $\sim 10.5$\,pc, we adopt the supernova ejecta mass $M_{\rm ej}=3\,M_\odot$, the explosion energy $E_{\rm SN}=3 \times 10^{50}$\,erg, and the mean ambient ISM density $n_{\rm ISM}=0.3\,$cm$^{-3}$. It follows that the velocity of the shock was $2.1 \times 10^3$ km s$^{-1}$ at the beginning time of the { ST} phase, $ t_{\rm ST} = 1.5$\,kyr, and is $6.4 \times 10^2$\,km\,s$^{-1}$  at the present time \cite[]{1999ApJS..120..299T}\footnote{See \citeauthor{2017AJ....153..239L}\ (2017) for a SNR calculator ``snrpy''}. 

Assuming simple dynamical evolution of the SNR in a homogeneous medium, the maximum energy of the accelerated particles in HB9 can be estimated following Equation (21) in \citet{2013MNRAS.431..415B} or Equation (27) in \citet{2015APh....69....1C}. The maximum energy was $\sim 20$\ TeV at the beginning of the { ST} phase and is $\sim 5$ TeV at present. These estimates can be appreciably lower if damping happens to play an important role. This may be the case especially at late times, when the non resonant hybrid instability can no longer be excited, and ion-neutral damping may be in action, making the maximum energy of the accelerated particles appreciably lower than estimated above { (see for instance Fig. 2 of \cite{ZP2003}).}
 
Since MC R1 is observed to have been reached by the SNR shock, the $\gamma$-ray emission from this location cannot provide information on the CR diffusion in the near-source region. Hence below we focus on MC R2, from which however at present there is no evidence of TeV $\gamma$-ray emission. It is not clear whether this may be due to a too low maximum energy or to the fact that the $>10$ TeV protons escape too quickly from the circumsource region.

\subsection{W28}

SNR W28, a prototypical thermal composite SNR, is one of the best-known middle-age remnants, located in a star-forming region which hosts HII regions M8 \citep[at a distance $\sim 2$ kpc,][]{2002ApJ...580..285T} and M20 \citep[at $\sim 1.7$ kpc,][]{1985ApJ...294..578L}. NANTEN $^{12}$CO\,($J$=1-0) data reveal that MCs HESS J1800$-$240\,A, B, and C to the south of the remnant appear coincident with the $\gamma$-ray emission from the TeV source HESS J1900$-$240 \citep{2008A&A...481..401A}. MC C is considered to be correlated with another SNR, G5.7$-$0.1 \citep[see e.g.,][]{2016ApJ...816...63J}, { and} hence we do not consider it here. 

We assume that they are at the same distance from the shock. The distance of W28 is suggested to be $\sim 1.9$ kpc based on HI observations \citep{2002AJ....124.2145V}, and its angular radius is $\sim 0.35^\circ$ ($\sim 12$ pc). Radio and near-infrared observations indicate that the ISM around W28 is made up of mainly low-density inter-clump medium which takes up to 90\% of the volume and dense clumps \citep{2005ApJ...618..297R}. The mean density of the ISM is $\sim 5$ cm$^{-3}$. The age of W28 is estimated to be $\sim 40$ kyr based on the cooling of the associated neutron star \citep{2004ARA&A..42..169Y}. We assume $E_{\rm SN} = 10^{51}$\,erg and $M_{\rm ej} = 6 M_\odot$. 
 
The SNR entered pressure-driven-snowplow (PDS) phase (in which all CRs are assumed to be released into the ISM) at time $t_{\rm PDS} \sim 5$ kyr after the explosion, corresponding to a radius of $\sim 7$ pc \citep{1988ApJ...334..252C,2017AJ....153..239L}. 
MCs A and B are located $\sim 0.45^\circ$ from the SNR center, corresponding to a distance $\sim 16$ pc. The masses of the two clouds are $6 \times 10^4 M_\odot$ and $4 \times 10^4 M_\odot$, respectively \citep{2008A&A...481..401A,2010sf2a.conf..313G}.

\section{Diffusion of CRs near SNRs}\label{sec:diffusion}

The transport of non-thermal particles in the near-source region is dominated by diffusion and advection, two processes that are properly described by the Parker equation:
\begin{equation}\label{eq:dif}
\frac{\partial f}{\partial t} + \nabla \cdot \left[\vec{u}(\vec{x},t)-D(p,\vec{x},t)\nabla f\right] - \frac{1}{3}\nabla \cdot \vec{u}(\vec{x},t)\frac{1}{p^2}\frac{\partial}{\partial p} \left[ p^3 f\right] = Q(p,\vec{x},t),
\end{equation}
where $f(p,\vec{x},t)$ is the isotropic part of the CR distribution function in phase space, $u(\vec{x},t)$ the advection velocity, and $D(p,\vec{x},t)$ the diffusion coefficient. The function $Q(p,\vec{x},t)$ describes the injection, in terms of particles leaving a SNR. In the following, we will illustrate our results in terms of two models for diffusion, one of phenomenological nature, in which $D$ is parametrized, and the other { in which $D$ is} calculated in a non-linear way, as a result of self-generation of waves. 

In both cases, the generation of $\gamma$-ray radiation is assumed to be due to the inelastic interactions of protons with the gas in a { MC} located at given distance from the SNR shock. In general, the flux of $\gamma$-rays of hadronic origin at photon energy $\varepsilon$ can be calculated as \citep{2021PhRvD.104l3027K}:
\begin{equation}\label{eq:L}
F_{\rm \gamma}(\varepsilon) = \varepsilon^2\frac{c}{4\pi d^2}\oint n_{\rm MC}(\vec{x}) \int_{\varepsilon/c}^\infty \frac{{\rm d} \sigma_\gamma}{{\rm d}\varepsilon}(\varepsilon,p)f(p,\vec{x},t)4\pi p^2{\rm d}p {\rm d} \vec{x}^3,
\end{equation}
where $n_{\rm MC}$ is the number density of the gas in the MC, ${\rm d} \sigma_\gamma(\varepsilon,p) /{\rm d}\varepsilon$ is the differential cross section of production of a photon at energy $\varepsilon$ by a proton with momentum $p$ \citep{2021PhRvD.104l3027K}.

\subsection{Phenomenological diffusion coefficient}

The phenomenological approach to the description of CR transport around the SNRs HB9 and W28 is based upon the assumption that the diffusion coefficient around these sources is spatially constant and numerically estimated to be a fraction of the Galactic diffusion coefficient $D_{\rm Gal}$, as estimated from the secondary/primary ratios \citep[see e.g.,][]{2009APh....31..284P}. The exact numerical value of $D_{\rm Gal}$ is not very important in this context, since it is only used as a reference with respect to which the reduced diffusivity manifests itself. Thus, we assume $D(p) = D_{\rm Gal}/\kappa$, with the suppression factor $\kappa>1$, $p$ the proton momentum, and for the Galactic diffusion coefficient we use $D_{\rm Gal}=3.6\times 10^{28}[{\rm pc}/(1\ {\rm GeV{ /c}})]^{1/3}$ { cm$^2$\,s$^{-1}$}, aware that deviations from this trend are likely, especially at low energies \cite[]{Evoli2020}. { This simple approach is adopted here following the same logic as in the past literature on the topic \cite[]{2008A&A...481..401A,2013MNRAS.429.1643N,Bao2019,Bao2021}, namely to have a quick and approximate assessment of the possible suppression of diffusivity in the near-source regions.}

The diffusion equation around the source in this phenomenological approach reads:
\begin{equation}\label{eq:difshell}
\frac{\partial f}{\partial t}(p,\vec{x},t) = D(p) \nabla^2 f(p,\vec{x},t) + Q(p,\vec{x},t),  
\end{equation}
where the $\nabla^2$ operator is written in 1D (i.e., the particles only diffuse along the $z$ axis) or 3D (i.e., the particles diffuse spherically symmetrically in the radial coordinate $r$) depending on the situation at hand and the spatial coordinate is $x=z$ in 1D and $x=r$ in 3D. The injection term $Q$ will be written as $Q_{\rm 1D}$ and $Q_{\rm 3D}$ in the two cases that we investigated{ , respectively}.

As discussed in \S \ref{sec:observations}, the age of W28 is such that at present it is expected to be in the PDS phase. Hence we assume that the release of the accelerated particles occurred impulsively at the beginning of the PDS phase ($t_{\rm PDS} \sim 5$ kyr for W28).  

For HB9, since it is still in its { ST} phase, we assume that the injection of particles occurs continuously in time but at a constant rate. In other words, we neglect the change in time of the velocity and { radius} of the forward shock. We do so to avoid making additional assumptions on the environment in which the explosion took place. 

The geometrical center of the problem, $\vec x=0$, is taken as the location of the injection of particles, so that $Q(p,\vec{x},t)$ can be written as 
\begin{equation}\label{eq:Q1D}
Q_{\rm 1D}(p,\rho,z,t_{\rm esc}) = \dot N_{\rm esc}(p,t_{\rm esc})\frac{H(\rho_{\rm Inj}-\rho)}{\pi \rho_{\rm Inj}^2}\delta(z) 
\end{equation}
in the 1D case and 
\begin{equation}
Q_{\rm 3D}(p,r,t_{\rm esc}) = \dot N_{\rm esc}(p,t_{\rm esc})\frac{\delta(r)}{4\pi r^2}. 
\end{equation}
in the 3D case, 
{ where $t_{\rm esc}$ is the time the particle escapes, $\rho$ the radius of the cylinder coordinate,} 
$\dot N_{\rm esc}(p,t_{\rm esc})$ the injection rate of protons, $H$ the Heaviside step function, and $\rho_{\rm Inj}$ the radial spread of the injection region (transverse size of the flux tube in the one dimensional case). We write the injection rate as
\begin{equation} \label{eq:Nesc}
\dot N_{\rm esc}(p,t_{\rm esc})=Q_{\rm SNR} \left(\frac{p}{\rm 1\ GeV/c}\right)^{-2-\alpha_{E}} H(p-{\rm 1\ GeV/c})e^{-p/p_{\rm cut}}\frac{H(t_{\rm esc}-t_{\rm ST})H(t_{\rm PDS}-t_{\rm esc})}{{\rm min}(t_{\rm age},t_{\rm PDS})-t_{\rm ST}},
\end{equation}
where $p_{\rm cut}$ is the cutoff momentum, and $Q_{\rm SNR}$ is the injection constant of the SNR which can be obtained using the normalization condition: 
\begin{equation}
\int_{0}^{t_{\rm age}} \int_{1\,{\rm GeV}/c}^\infty 4\pi p^2 (\sqrt{c^2p^2+m_{\rm p}^2c^4}-m_{\rm p}c^2) \dot N_{\rm esc}(p,t_{\rm esc})\,{\rm d}p\,{\rm d}t_{\rm esc} = \xi E_{\rm SN}.
\end{equation}
Here $\xi$ is the efficiency of conversion of the SNR kinetic energy to CR energy (typically $\xi \sim 0.1$). 
In the 1D case, the choice of the SNR radius at the time of particle escape ($\rho_{\rm Inj}$) is influenced by several uncertainties: it depends upon the the detailed gas distribution in the region around the SNR and the type of supernova event we deal with. Technically, the uncertainties are consolidated into { chosen} different values of $\rho_{\rm Inj}$, and the implications of the different choices will be discussed. 

{ As discussed above, it is possible that escape of accelerated particles occurs only from the upstream region of the shock. In this case the spectrum of the escaping particles may be limited to a narrow neighborhood of the maximum energy at that given time, $E_{max}(t)$, while the spectrum in the near-source region from the overlap of previous times is expected to show a low energy cut at $E_{max}(t)$. Occasionally in the calculations that follow we will adopt this recipe to mimic the scenario in which escape occurs in this way. }

For HB9, which is currently in the { ST} phase, we calculate the mass of MCs based on an average { tube} radius of { $R_{\rm tube} = 8$\,pc}. This average is determined from its present radius of 10 pc and its initial radius of 5 pc at the start of the { ST} phase. { We adopt these values as benchmark while we will warn the reader when other possibilities will be considered.}

In contrast, W28, due to its high mean ISM density of approximately 5\,cm$^{-3}$\citep{2005ApJ...618..297R}, enters its PDS phase early at about 5 kyr and has a shock radius of roughly 7 pc. Given its current age of $\sim 40$ kyr, we assume that all particles escaped at the same time $t_{\rm PDS}$ when the radius of the SNR was $\rho_{\rm Inj}=7$ pc.

The maximum energy of particles accelerated at the shocks in the two selected SNRs depends on numerous ingredients that are poorly known (such as possibility to excite the non-resonant instability at some time during the evolution of the remnant, role of damping and conditions in the circum-source region). We check that it is likely that there were the conditions for the excitation of the non-resonant instability at the beginning of the { ST} phase, though for some choices of the parameters the condition is satisfied only marginally. For the sake of the present calculations, given the large uncertainties in the maximum energy and its dependence on time, we assume here that the spectrum of the CRs has a fixed exponential cutoff at 20 TeV for HB9 and 100 TeV for W28. 

In the 1D situation, the diffusion problem has a cylindrical symmetry and if the diffusion coefficient is assumed to be spatially constant, the solution can be written as 
\begin{equation}\label{eq:dif1d}
f(p,z,t_{\rm age}) =\int_{0}^{t_{\rm age}} \frac{\dot N_{\rm esc}(p,t_{\rm esc})}{\pi^{1/2}R_{\rm d}}\frac{H(\rho_{\rm Inj}-\rho)}{\pi \rho_{\rm Inj}^2} e^{-z^2/R_{\rm d}^2} {\rm d}t_{\rm esc},
\end{equation}
where $R_{\rm d} = \sqrt{4D_{\rm ISM}(t_{\rm age}-t_{\rm esc})}$ is the distance that particles can diffuse away from the source at the given time $t_{\rm age}$.

\subsection{Self-generated diffusion}\label{sec:model}

In the region surrounding a source we can naturally expect CR energy density in excess of the mean Galactic value, as well as stronger CR gradients that result from the presence of a source. In these conditions, as previously discussed by \citep{2008AdSpR..42..486P,Malkov2013,2016PhRvD..94h3003D,2016MNRAS.461.3552N,2018MNRAS.474.1944D,2019MNRAS.484.2684N,2022A&A...660A..57R}, the transport of CRs may be dominated by self-generated perturbations, mainly in the form of Alfv\'en waves moving parallel to the regular Galactic magnetic field in which the source is assumed to be located. Since the gradient has a well defined sign, most Alfv\'en waves move away from the source. This results in an advective term in the transport equation with an advection velocity that equals the Alfv\'en speed. The main novelty of the present calculations is the fact that we account for the possible presence of dense MCs { located in the regions around specific sources and with properties that may lead to the identification of reduced diffusivity. Moreover we investigate the effect of the MCs} in terms of grammage accumulated by CRs while propagating away from the sources. 

It is important to realize that the process of self-generation is intrinsically non-linear: the level of perturbations that is induced by this phenomenon is related to the local density of CRs at a given location, but on the other hand the CR density is larger when the diffusion coefficient is smaller, namely when the self-generation is more effective. The rate of particle injection per unit volume, and hence the spatial extent of the source region are also important parameters of the problem. 

In the { below} we will assume that the transport of CRs in a region of size { similar to the coherence length $L_{\rm c}$} around the source can be considered approximately one dimensional \citep{2008AdSpR..42..486P}. 
{ Following \cite{2016JCAP...05..056B}, we adopt a value $L_{\rm c} \approx 220$\,pc for the Galactic magnetic field $\vec{B}_0$.}
When the particles reach a distance from the source that is appreciably larger than $L_{\rm c}$, the transport is expected to change over to a 3D phenomenon \citep{2013PhRvD..88b3010G}. When this happens the CR density and its gradient become smaller and the role of self-generation quickly becomes unimportant in terms of confinement on regions of size $L_{\rm c}$. To be more precise, we should say that the local gradients become unimportant, while it is possible that the gradient on Galactic scales may drive the turbulence responsible for the transport of the bulk of CRs on Galactic scales \cite[]{Blasi2012}, at least up to energies $\lesssim$TeV.

As mentioned above, the calculations are carried out in the so-called flux tube approximation, namely we assume that the CRs remain trapped in a region whose cross section is the transverse size of the source that injects them into the ISM. Recently, hybrid particle-in-cell simulations \cite[]{2021ApJ...914L..13S,2022MNRAS.512..233S} were used to show that the reduced diffusion inside the flux tube results in an enhanced pressure gradient that leads the tube to inflate. We neglect this aspect here since it is expected to be the most important for young SNRs. 

The 1D diffusion-advection equation reads
\begin{equation}\label{eq:dif2}
\frac{\partial f(p,z,t)}{\partial t} + u\frac{\partial f}{\partial z} - \frac{\partial}{\partial z}\left[D(p,z,t)\frac{\partial f}{\partial z}\right] - \frac{{\rm d}u}{{\rm d}z}\frac{p}{3}\frac{\partial f}{\partial p} = Q(p,z,t).
\end{equation}
Since in weakly perturbed fields $u$ is $+v_{\rm A}$ { ($v_{\rm A}:=B_0/\sqrt{4\pi m_{\rm i}n_{\rm i}}$)} for $z>0$ and $-v_{\rm A}$ for $z<0$, d$u/$d$z = 2v_{\rm A}\delta(z)$. The diffusion coefficient $D(p,z,t)$ can be calculated as
\begin{equation}\label{eq:D}
D(p,z,t) = \frac{r_{\rm L}(p)c}{3\mathcal{F}(k,z,t)|_{k = 1/r_{\rm L}(p)}},
\end{equation}
where $r_{\rm L}$ is the Larmor radius of the particles and the dimensionless spectral power per unit $\ln(k)$, $\mathcal{F}(k,z,t)$, is calculated at the resonant wave number $k = 1/r_{\rm L}(p)$.

The evolution of $\mathcal{F}(k,z,t)$ is described by the equation:
\begin{equation}\label{eq:wave}
\frac{\partial\mathcal{F}}{\partial t} + u\frac{\partial \mathcal{F}}{\partial z} = (\Gamma_{\rm CR}-\Gamma_{\rm D})\mathcal{F}(k,z,t),
\end{equation}
where the growth rate of self-generated turbulence is given by \cite[]{1971ApJ...170..265S}
\begin{equation}\label{eq:gr}
 \Gamma_{\rm CR} = \frac{16\pi^2v_{\rm A}}{3\mathcal{F}B_0^2}\left|p^4c\frac{\partial f}{\partial z}\right|_{p=eB_0/(kc)},
\end{equation}
and $\Gamma_{\rm D}=\Gamma_{\rm NLL}+\Gamma_{\rm FG}+\Gamma_{\rm IND}$ is the damping term, where $\Gamma_{\rm FG} = 2kv_{\rm A}/\sqrt{kL_{\rm c}}$ is the damping rate from interactions with pre-existing MHD turbulence \citep{2004ApJ...604..671F}, 
\begin{equation}
\Gamma_{\rm NLL} = \sqrt{\frac{\pi}{2}\frac{k_{\rm B}T}{m_{\rm p}}} \frac{\mathcal{F}}{r_{\rm L}}
\end{equation}
is the rate of non-linear Landau damping \citep[see e.g.,][and references therein]{2013ApJ...767...87W,2019MNRAS.484.2684N}, $\Gamma_{\rm IND}$ is the ion-neutral damping rate, and $k_{\rm B}$ the Boltzmann constant. The $\Gamma_{\rm IND}$ can be obtained by solving the equation \citep{1982ApJ...259..859Z}
\begin{equation}\label{eq:IND0}
\omega(\omega^2-\omega_k^2)+i\nu \left[(1+\epsilon)\omega^2-\epsilon\omega_k^2 \right]=0,
\end{equation} 
where $\omega_k = kB_0/\sqrt{4\pi n_{\rm i}m_{\rm i}}$, { $\omega$ is the complex frequency of the wave,} $\nu$ { the} ion-neutral momentum transfer ratio 
\begin{equation}
\nu \approx 1.68 \times 10^{-8}\left(\frac{n_{\rm n}}{{\rm cm}^{-3}}\right) \left(\frac{T}{10^4{\rm\,K}}\right)^{0.4},
\end{equation}
and $\epsilon=n_{\rm i}/n_{\rm n}$ { is the ratio of ion density to neutral density}. With $\omega = \omega_{\rm R} + i\Gamma_{\rm IND}/2$, \autoref{eq:IND0} can be written as 
\begin{equation}
\omega_{k}^2 = -\frac{\Gamma_{\rm IND}}{\Gamma_{\rm IND}+\nu}\left[\Gamma_{\rm IND}+\nu(1+\epsilon)\right]^2.    
\end{equation}

The equation{ s} are solved using boundary conditions that $\mathcal{F}|_{z=L_{\rm c}/2} = \mathcal{F}_{\rm Gal}$ and a free escape boundary condition $f|_{z=L_{\rm c}/2} = 0$ for the particles, where $\mathcal{F}_{\rm Gal} = r_{\rm L}c/[3D_{\rm Gal}(E)]$ is the Galactic turbulence spectral power. The injection term $Q(p,z,t)$ is the one described in Equations \ref{eq:Q1D}.

\section{Results}\label{sec:Results}

Here we illustrate the results of our calculations of CR transport around sources using phenomenological models and self-generation models and their implications in terms of $\gamma$-ray emission from { the} MCs located in the region{ s} around two SNRs, HB9 and W28. 

\subsection{Phenomenological models of CR transport}\label{sec:pheno}

The phenomenological models allow us to infer some generic properties of CR transport in the near-source regions without delving with the complex aspects involved in self-generation. In particular, these models can provide an easy flag for appreciable suppression of diffusivity in the near-source regions { (see for instance \cite{2007Ap&SS.309..365G,2013MNRAS.429.1643N})}.

\begin{center}
\begin{figure}
\includegraphics[scale=0.45]{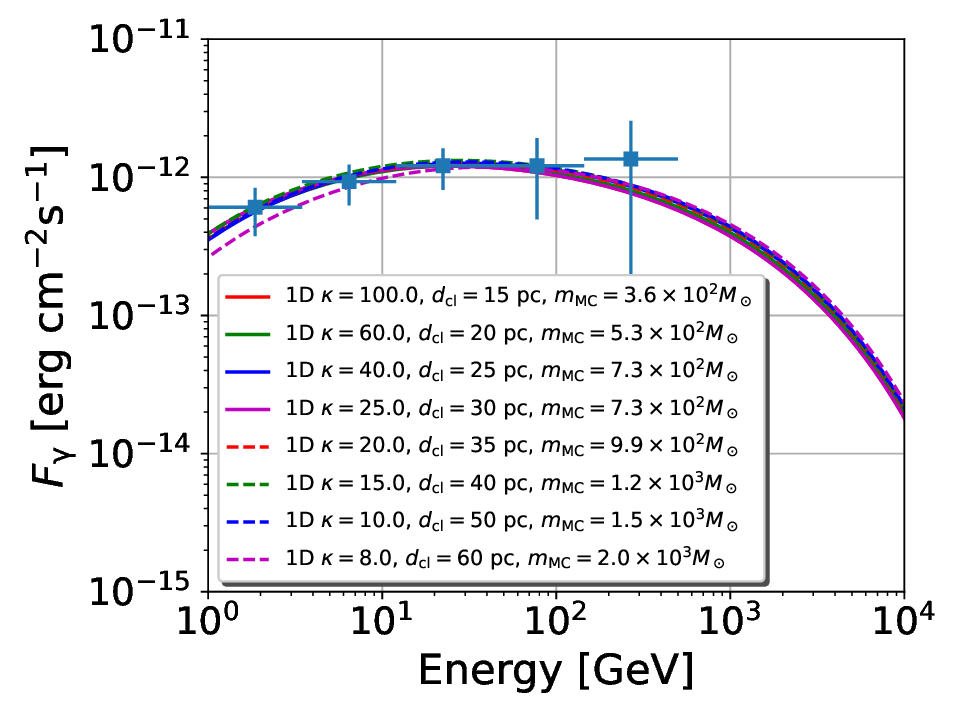}
\includegraphics[scale=0.45]{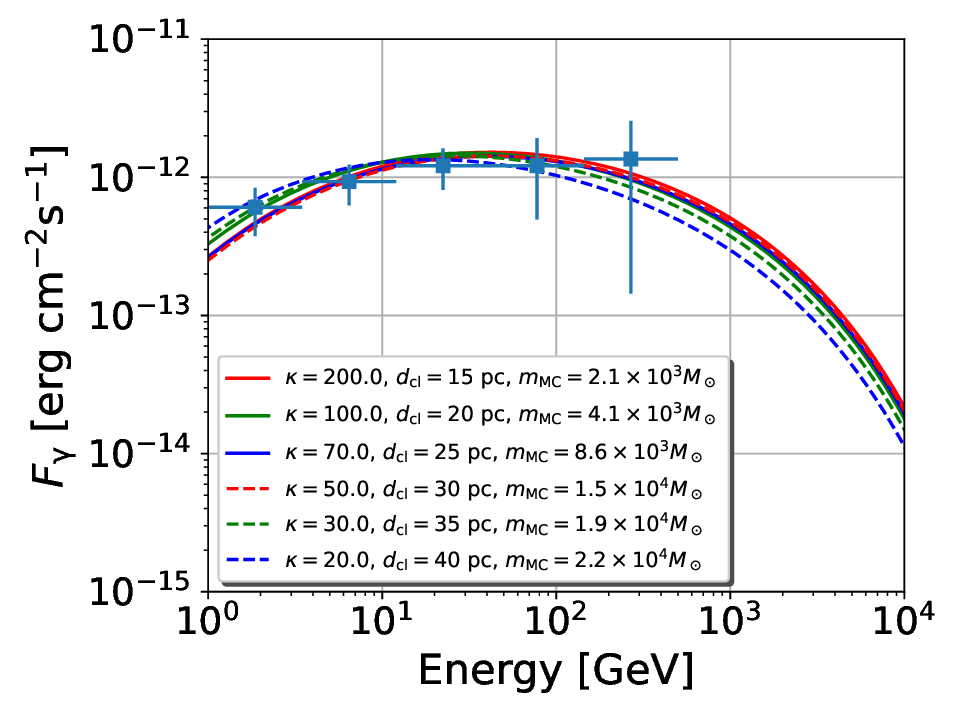}
	\caption{{ H}adronic $\gamma$-ray emission from MC R2 around HB9 for the 1D (left panel) and 3D (right panel) phenomenological models. The different curves refer to different { combinations of values of the suppression factor for the diffusivity, the distance of the cloud from the SNR, and the mass of MC}. { $\xi = 9\%$ is adopted in all panels.}}
\label{fig:Phen}
\end{figure}
\end{center}

The $\gamma$-ray emission expected from the MC (R2) near the HB9 SNR is shown in \autoref{fig:Phen}, for the 1D (left panel) and 3D (right panel) case, respectively. The different curves refer to different distances and masses of the R2 MC, so as to illustrate the effect of the uncertainty in these two quantities. One can see that for the ranges of mass and distance that are allowed by observations, a suppression of the diffusion coefficient with respect to the Galactic one is required, for both the 1D and 3D cases. The suppression factor $\kappa$ ranges between $\sim 10$ and $\sim 200$. 

{ It is important to stress that} the main need for a suppression in the diffusion coefficient comes from the spectral shape of the $\gamma$-ray emission in the low energy range: the downward inflection of the spectrum requires that the low energy particles do not reach the cloud. Clearly this effect is more pronounced, namely requires more suppression of the diffusion coefficient, when the distance from the SNR shock to the cloud $d_{\rm cl}$ is larger.

{ These results would suggest that for the HB9 SNR there is evidence of suppressed diffusivity and that such evidence is independent of the assumed dimensionality of the problem. Given the importance of the statement, it is important to check whether other factors may mimic the same phenomenology. One possibility is related to the way particles escape the remnant: as discussed above, the general, and somewhat idealized, picture of particle escape is that CRs leave the remnant from upstream of the shock in such a way that only particles with energy in a narrow neighborhood of the maximum energy at that given time can escape. This means that at a given time only particles with energy $>E_{max}(t)$ can fill the medium surrounding the SNR. From the phenomenological point of view we can mimic this picture by repeating the previous calculation but assuming that the spectrum of escaping particles has a minimum energy and check how low such minimum energy should be in order to fit the observed $\gamma$-ray emission.}

\begin{center}
\begin{figure}
\includegraphics[scale=0.65]{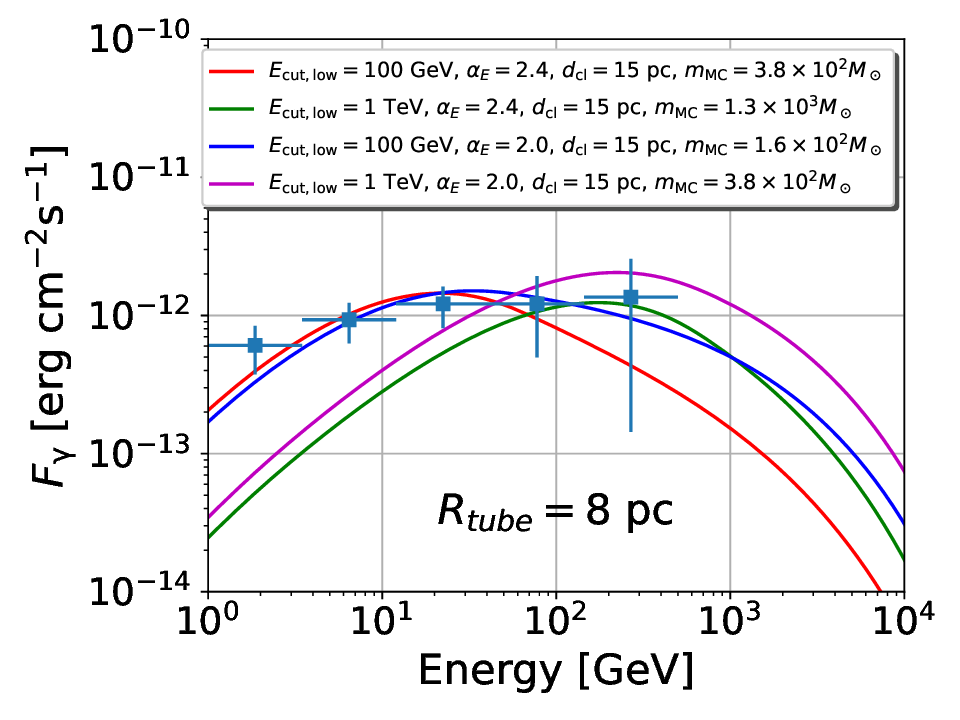} 
	\caption{{ H}adronic $\gamma$-ray emission from MC R2 around HB9 for the 1D phenomenological model, assuming a minimum energy of the escaping particles as indicated in the legend. Two injection spectra have been adopted, with slope{ s} $\alpha_E=2$ and $2.4$.}
\label{fig:LowEmin}
\end{figure}
\end{center}

{ The results of this exercise are summarized in Fig. \ref{fig:LowEmin}, where we allowed the minimum energy to be $1$ TeV and $100$ GeV and the diffusion coefficient is the same as assumed for the Galaxy at large. The suppression of the $\gamma$-ray emission at low energies here is not due to confinement but rather to the fact that only particles with energies above some minimum value have escaped the remnant at the time corresponding to the age of HB9. The low energy behaviour of the $\gamma$-ray emission reflects the energy dependence of the differential cross section of pion production $\propto E_{\pi}^{-1}$. 

This exercise, based on a phenomenological description of the diffusion of particles in the near-source region, suggests that there is a degeneracy between the model of particle escape and the phenomenon of reduced diffusivity: the observed $\gamma$-ray emission is both compatible with a maximum energy in the hundred GeV range at the age of HB9 or with the escape of all particles from the shock and a suppressed diffusivity around the remnant, although the physical reason for the low energy suppression in the $\gamma$-ray emission, as discussed above, is completely different in the two scenarios. 

The issue of whether it is reasonable to have a maximum energy in the $\sim 100$ GeV range for a SNR having the age of HB9 is strongly related to the damping mechanisms that are at work in the acceleration region. For instance, in Fig. 2 of \cite{ZP2003} one can read the maximum energy as a function of time in the presence of damping. If only non-linear Landau damping is at work, it is not feasible to decrease the maximum energy to such low values. If ion-neutral damping near the shock dominates, a maximum energy below $\sim 1$ TeV seems somewhat too low for HB9, but it is hard to rule it out, given the uncertainties on the local density and the ionization fraction in the acceleration region. Hence, unfortunately some level of degeneracy between the two scenarios discussed above cannot be excluded. 
}

\begin{center}
\begin{figure}
\includegraphics[scale=0.45]{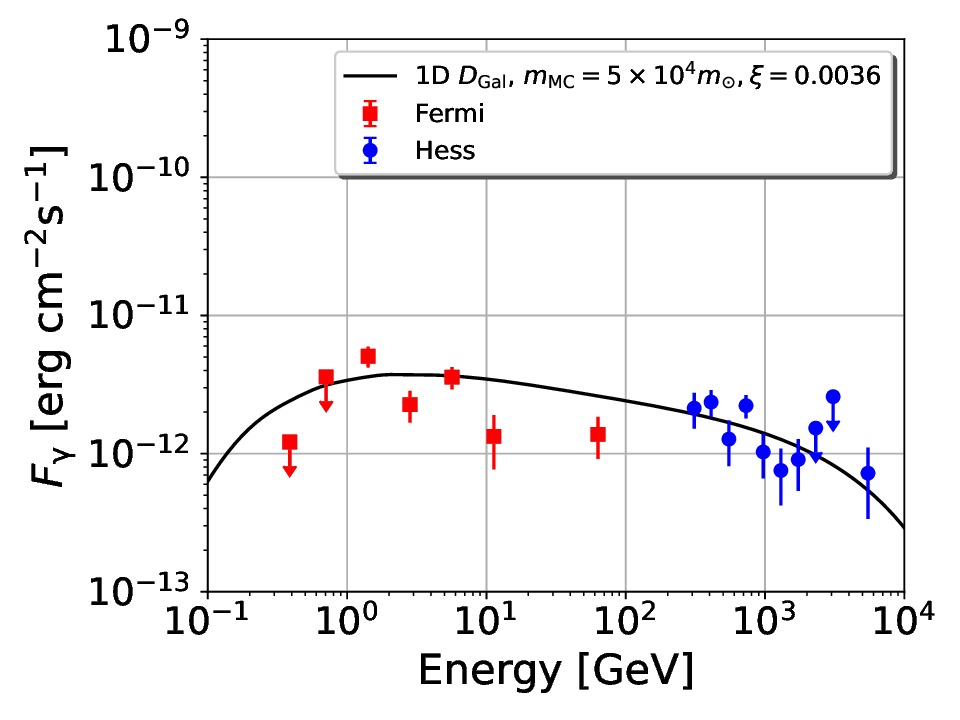}
\includegraphics[scale=0.45]{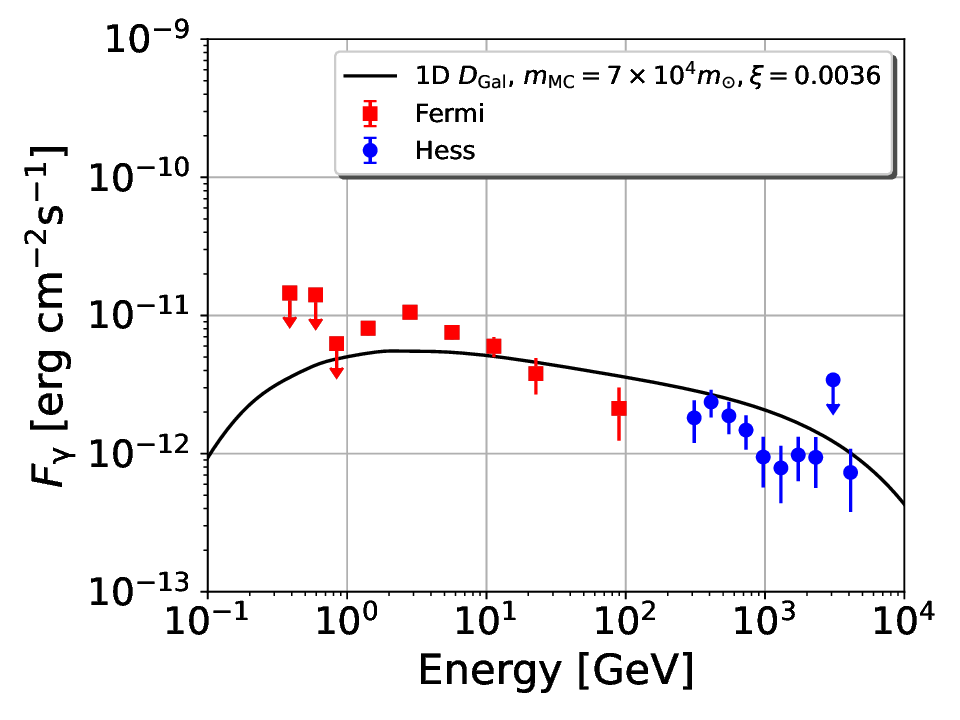}
\\
\includegraphics[scale=0.45]{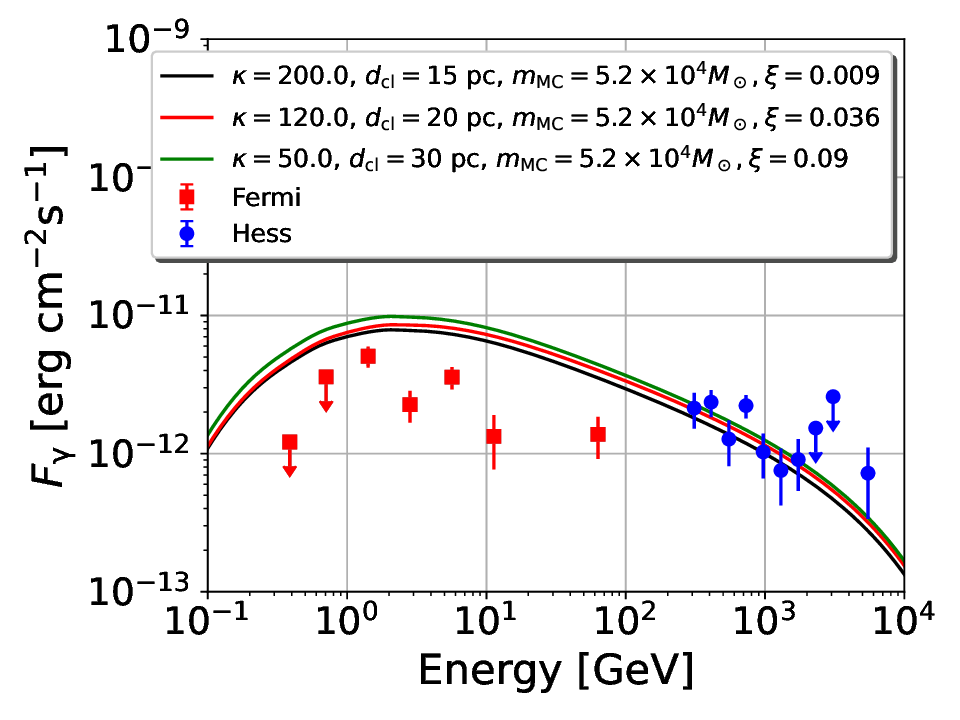}
\includegraphics[scale=0.45]{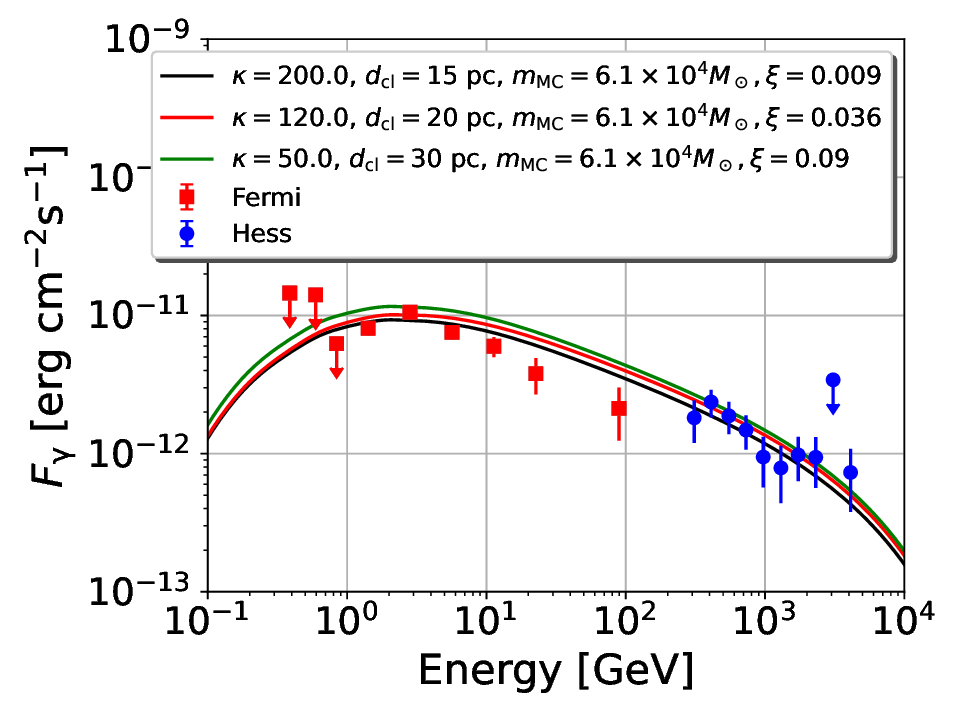}
	\caption{Phenomenological models for the hadronic $\gamma$-ray emission from MC A (left panels) and B (right panels) of the W28 complex. The different curves refer to different values of the suppression factor for the diffusivity, different value of the distance of the cloud from the SNR and different masses of the MC. The top panels illustrate the situation in the 1D models, while the 3D cases are shown in the bottom panels.}
\label{fig:PhenW28}
\end{figure}
\end{center}

The $\gamma$-ray emission expected in our phenomenological models of CR transport around the W28 SNR is shown in the left (right) panel of \autoref{fig:PhenW28} for MC A (B). Our predictions are compared with the $\gamma$-ray observations of Fermi-LAT and H.E.S.S. We use the nominal values of the mass of the two MCs.

In the 1D models, one can see that the observations are well described by adopting the Galactic diffusion coefficient with no need for suppression. Note that the acceleration efficiency $\xi$ required in these calculations is extremely low, of order 0.3\%. {We also find that in the 1D situation, $d_{\rm cl}$ can hardly affect the spectra because for $z \ll R_{\rm d} \sim 4.0\times 10^2 (E/1\ {\rm TeV})^{1/6}$ pc, the distribution function $f(p,z,t) \propto 1/R_{\rm d}$, but is independent on $d_{\rm cl}$.} This might suggest that not the whole volume of the clouds is actually illuminated by the CRs accelerated in W28, so that more reasonable values of the efficiency can be adopted. In any case, larger and more common values of the acceleration efficiency would strengthen the evidence that no suppression in the diffusion coefficient is required in the case of 1D transport around W28, consistent with the earlier results by \cite{2013MNRAS.429.1643N}.

In the 3D case of CR transport, as expected, the requirement for suppressed diffusivity becomes evident: for values of the distance $d_{\rm cl}$ between the cloud and the remnant that are accepted in the literature and for the nominal values of the masses of the two clouds, the diffusion coefficient is required to be suppressed by a factor $\sim 50-200$. This is a generic feature due to the fact that in the 3D case the CR density drops faster with the distance and hence a smaller diffusion coefficient is required to obtain the same $\gamma$-ray emission. 

{ As discussed earlier in this manuscript, this latter scenario is of limited interest in that, in order for the diffusion to be treated as three dimensional, the coherence scale of the turbulent field is required to be much smaller than the size of the SNR, which by itself suggests that the diffusion coefficient is smaller than the Galactic value. Notice also that, from a purely observational point of view, the data on the low energy $\gamma$-ray emission from W28 are not unambiguously suggestive of a flux reduction, namely there is no clear need for a suppression of diffusivity.}

\subsection{Models of self-generated transport}

The excitation of streaming instability near the source leads to a reduction of the diffusion coefficient that in turn increases the CR density (CR confinement). The solution of the transport equation is made difficult by the fact that the diffusion coefficient entering the diffusion equation depends on the distribution function of particles, making the problem non-linear. Moreover, the spectrum of self-generated perturbations evolves in time trying to reach a balance between the excitation of the instability (growth) and damping. The latter depends rather critically on the level of ionization of the background plasma. 

Below we illustrate the effects of this phenomenon in terms of the CR confinement and $\gamma$-ray emission in the regions around HB9 and W28. We focus on a 1D problem in which CRs propagate non-linearly along the local flux tube determined by the direction of the Galactic magnetic field in the region where the source is located. The 3D case is less interesting in terms of self-generation, since the small coherence scale of the field necessary to justify the assumption of 3D transport already implies that the local diffusion coefficient is smaller than in the Galaxy at large. 

For the sake of simplicity and to limit the number of parameters, we assume that the transverse sizes of the SNR and of the cloud are the same ($\rho_{\rm Inj}$) and that the cloud is a cylinder extending from a distance $z_1$ to a distance $z_2$ away from the source (see \autoref{tab:par} for the numerical values of these parameters { as well as} the cloud mass). 

\begin{center}
\begin{figure}

\includegraphics[scale=0.5]{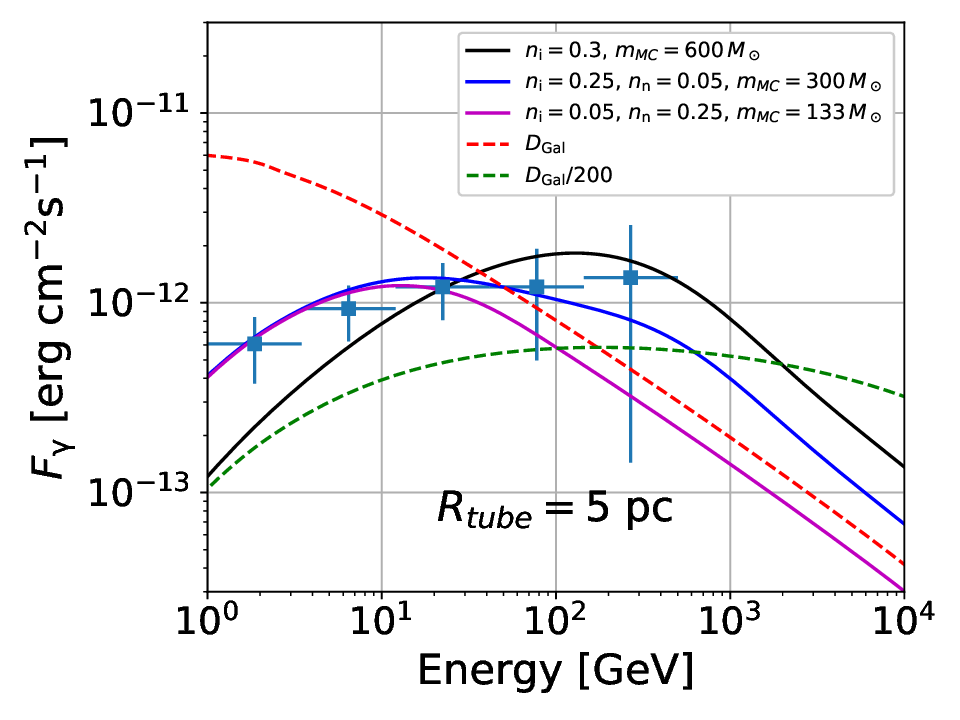}
\includegraphics[scale=0.5]{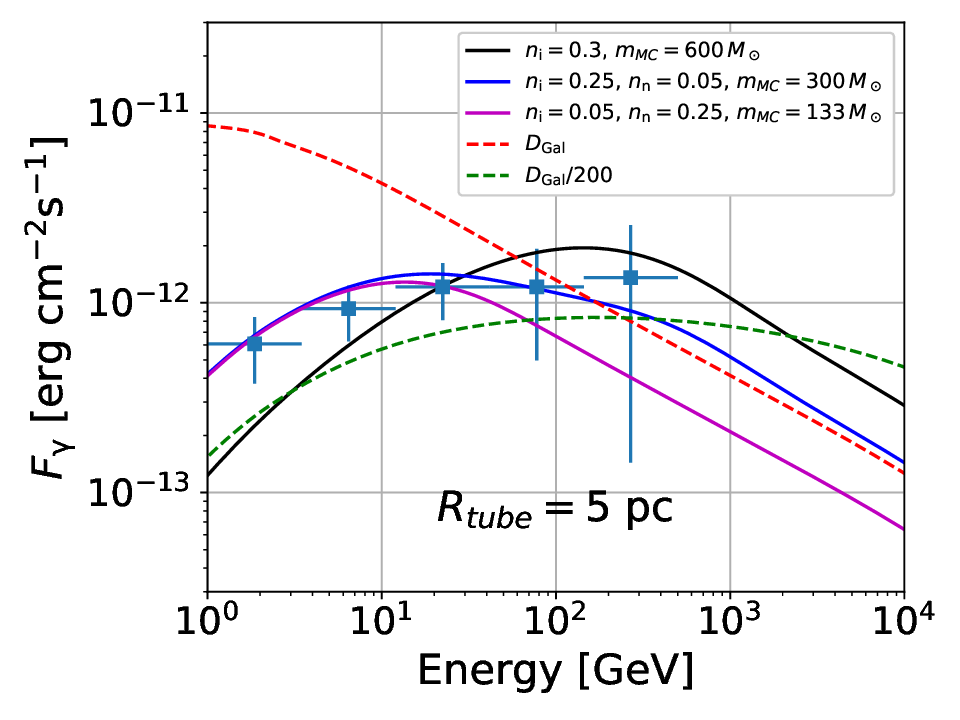}\\
\includegraphics[scale=0.5]{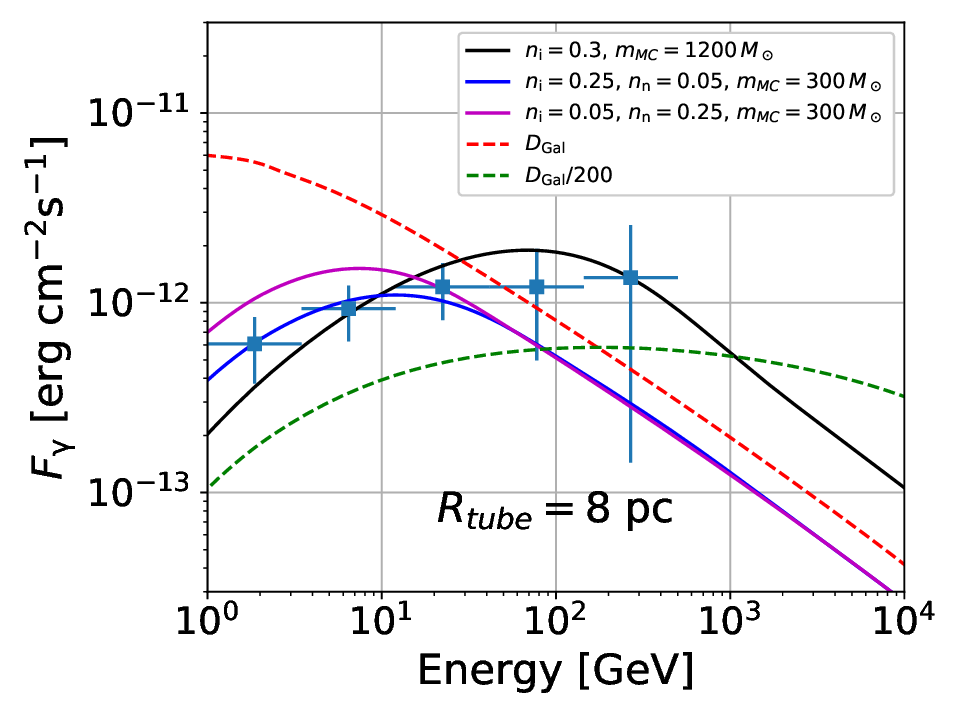}
\includegraphics[scale=0.5]{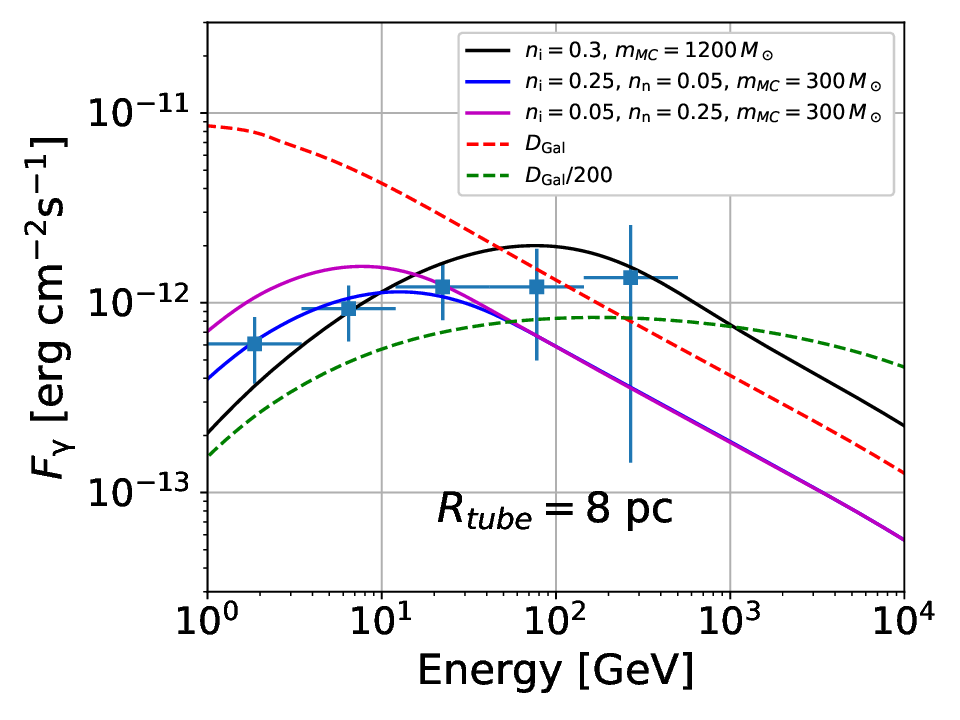}\\
\includegraphics[scale=0.5]{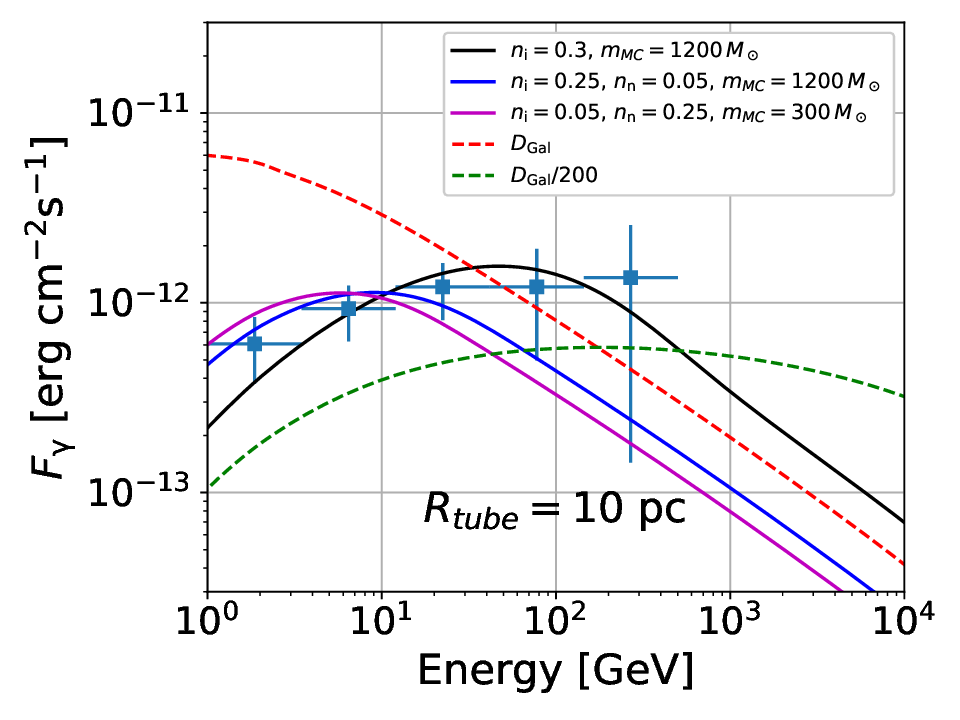}
\includegraphics[scale=0.5]{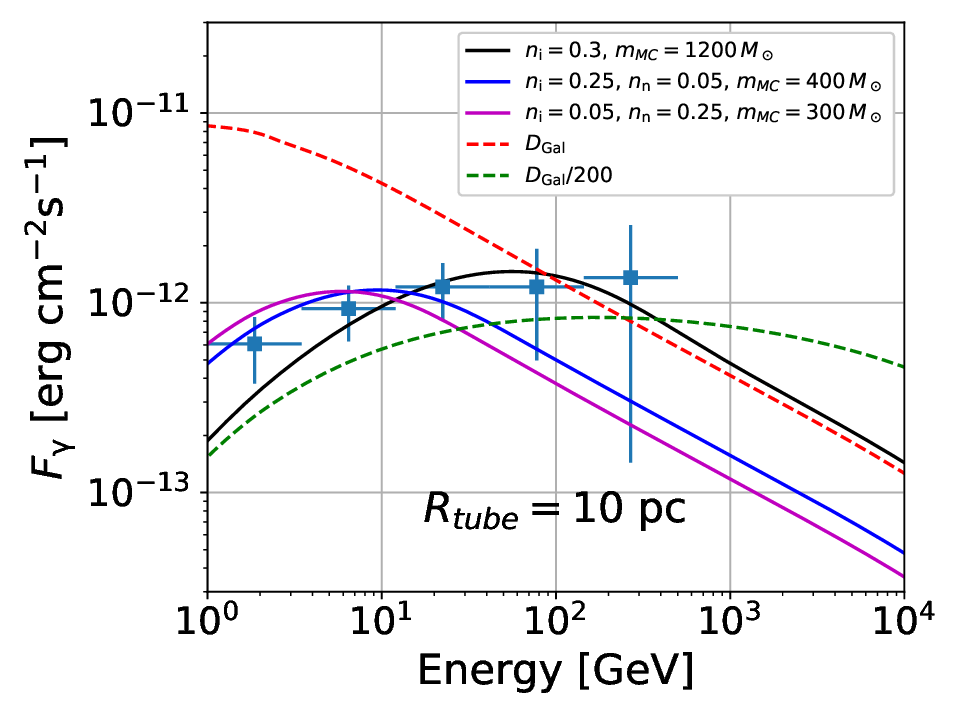}
\caption{$\gamma$-ray emission from { MC} R2 { near HB9} in the self-generated turbulence case. In all curves we have considered the convection term. In the left panel we assume the free escape boundary at 100 pc while in the right panel we assume the free escape boundary at 300 pc. { $\xi = 7\%$ is adopted in all panels. The black, blue and magenta lines show the spectra resulting from our calculations with $(n_{\rm i}, n_{\rm n})=(0.3,0)$, (0.25, 0.05), (0.05, 0.25) $\rm cm^{-3}$, respectively. The red dashed lines represent the $\gamma$-ray spectra with $D=D_{\rm Gal}$, while the the green dashed lines refer to the case $D=D_{\rm Gal}/200$.}}
\label{fig:R1R2}
\end{figure}
\end{center}

\begin{center}
\begin{figure}
\includegraphics[scale=0.65]{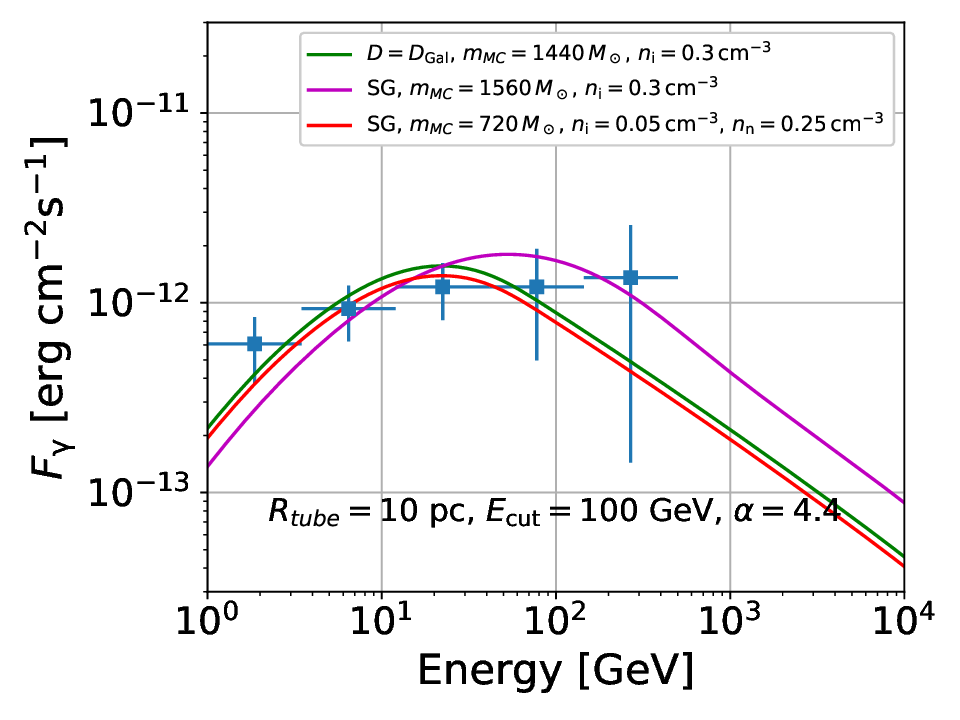}
\caption{ $\gamma$-ray emission from MC R2 near HB9 in the case of a spectrum of particles with a low energy cut at $E<100$ GeV. The three curves show the case of Galactic diffusion coefficient, the case of self-generation with $R_{\rm tube}=10$\,pc, $(n_{\rm i}, n_{\rm n})=(0.3,0)$ $cm^{-3}$ and the case of self-generation with $R_{\rm tube}=10$\,pc, $(n_{\rm i}, n_{\rm n})=(0.05,0.25)$ $cm^{-3}$. The power-law index in the momentum space $\alpha=4.4$ is used.}
\label{fig:lowEminNL}
\end{figure}
\end{center}

The spectrum of $\gamma$-ray emission from MC R2 around HB9 in the self-generated case is shown in \autoref{fig:R1R2} for the case of fully ionized and partially ionized plasma in the { near-source} region. The left (right) panel refers to a coherence length $L_{\rm c}=100$ pc (300 pc). The injection spectrum adopted for the computation is $\propto E^{-2.4}$ { all the way down to small energies. As discussed in Sec. \ref{sec:pheno}, it is possible that this is not the case and that escape from upstream of the shock is concentrated in a narrow energy region around the maximum energy at that given time. This picture results in an effective low energy cutoff in the spectrum injected over time, at the maximum energy reached at the age of HB9. We will comment below on how this scenario affects self-generation.} 

In the partially ionized case, ion-neutral damping leads to substantial suppression of Alfv\'en waves produced at high wavenumbers, responsible for resonances with low energy CRs. As a consequence, the diffusion coefficient of the CRs with energies $\lesssim 1$ TeV becomes large and inhibits self-confinement. In the case in which the gas is fully ionized, the main channel of damping is non-linear Landau damping, but its effect is not sufficient to prevent self-confinement. This can be seen very clearly in \autoref{fig:R1R2}, which shows a downturn of the $\gamma$-ray emission at energies $\lesssim 100$ GeV, corresponding to the fact that the diffusion coefficient is lower and the low energy particles can hardly reach the R2 MC. This interpretation can be easily validated by looking at the $\gamma$-ray emission obtained for a Galactic diffusion coefficient (red curve{ s}) and a Galactic diffusion coefficient suppressed by hand by a factor 200 (green curve{ s}). The latter shows a similar downturn in the $\gamma$-ray emission at energies $\lesssim 100$ GeV. The main difference between the cases with $L_c=100$ pc and $L_c=300$ pc is that in the latter the confinement extends to higher energies, as expected. 




{ Taken at face value, this finding would be clear evidence that even in the case of an under-luminous SNR such as HB9 there is suppression of diffusivity around the source. Unfortunately, as discussed in Sec. \ref{sec:pheno}, the low energy $\gamma$-ray flux reduction suggesting a smaller diffusivity can also be caused by the differential escape of CRs from the upstream region of the shock, if the maximum energy of the accelerated particles for a SNR with the age of HB9 is as low as $\sim 100$ GeV. Although this condition requires a role of ion-neutral damping at the shock that is probably larger than what one would expect \cite[]{ZP2003}, the scenario cannot be ruled out given the uncertainties in the environmental parameters.

In order to show more clearly the effect of this degeneracy, in Fig. \ref{fig:lowEminNL} we show three cases of CR transport in a situation in which the spectrum of escaping particles is characterized by a low energy cut at $100$ GeV: the first case corresponds to assuming that the diffusion coefficient is the same as in the Galaxy at large; in the second case the self-generation is active but there is no neutral gas; the third case shows the result of self-generation with ion-neutral damping taken into account. The plot clearly shows that the observed $\gamma$-ray emission simply requires a small renormalization of the cloud mass in the three cases, within the uncertainties on such parameter. Hence the evidence for self-generation around HB9 can, at present, only be considered as circumstantial.   
}

\begin{figure}
\includegraphics[scale=0.45]{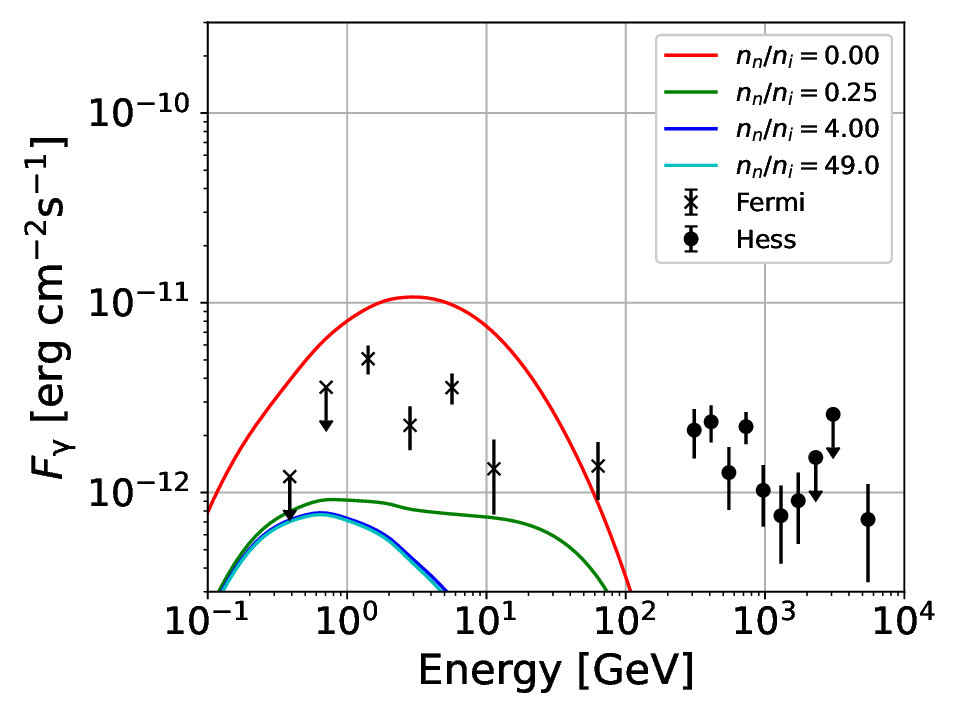}
\includegraphics[scale=0.45]{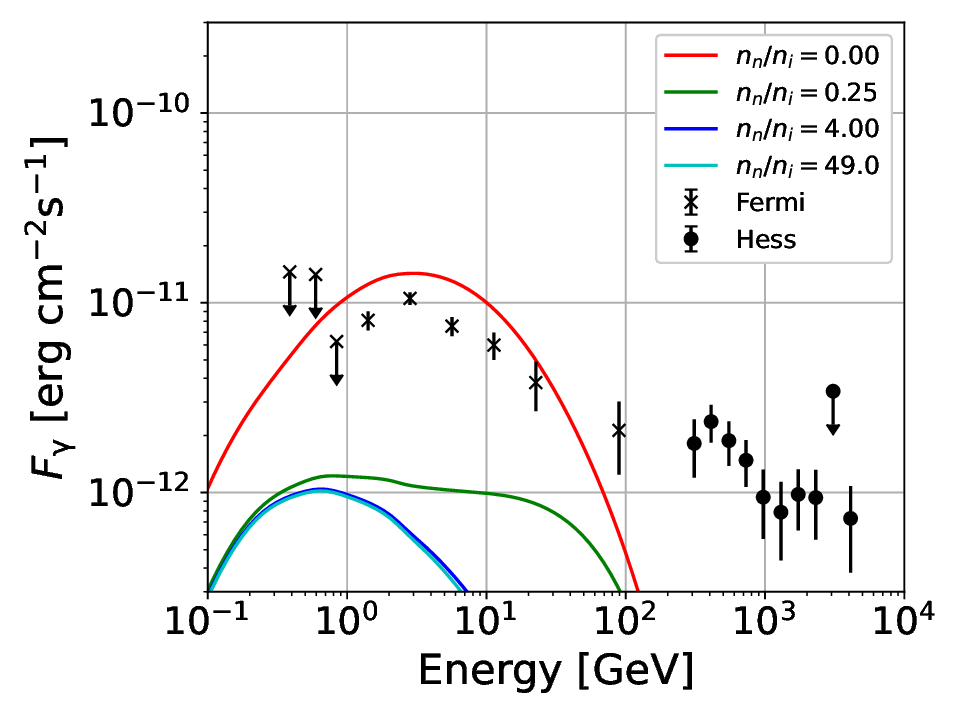}\\
\includegraphics[scale=0.45]{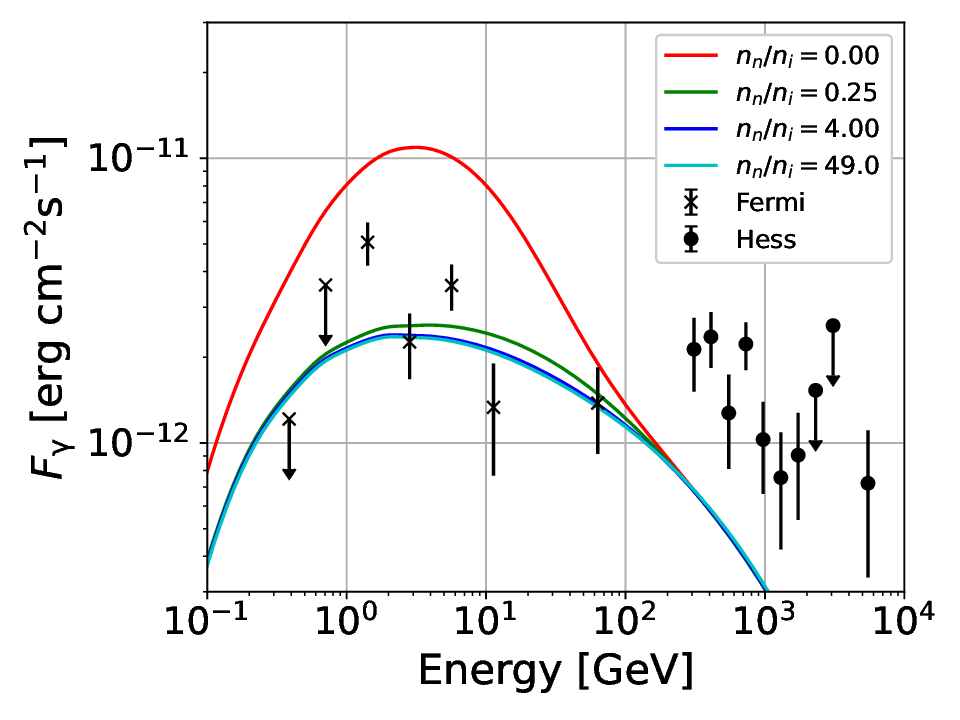}
\includegraphics[scale=0.45]{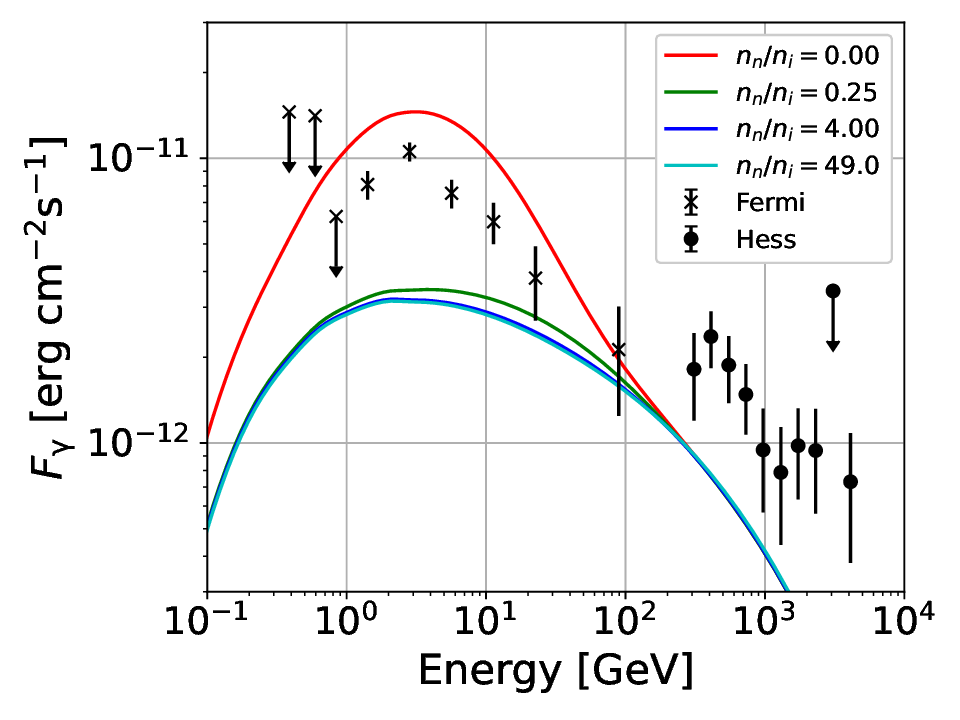}

\caption{$\gamma$-ray emission from MC A (left panels) and MC B (right panels), modelled in the self-generated turbulence case. For all { curves} we have considered the convection. The injection term is impulsive in time at the beginning of { ST} phase. In the first row we assume the free escape boundary at 100 pc, in the second row we assume the free escape boundary at 300 pc. In all rows we assume { that} the total proton energy is $4\times 10^{48}$ erg.}
\label{fig:IND}
\end{figure}

In \autoref{fig:IND}, we show the $\gamma$-ray fluxes from MC A (left panels) and MC B (right panels) around W28 for different values of the $n_n/n_i$ { ratio}, where the total density is fixed to $5\rm cm^{-3}$. One can find that the effect of self-generation is only visible in the unlikely case of high density of fully ionized gas (red curve), { and the resulting $\gamma$-ray emission is not well described by this situation. In} all cases with neutral gas the ion-neutral damping is so severe that the diffusion coefficient at all proton energies of relevance here remains close to the Galactic diffusion coefficient. In fact, the low energy suppression in all the curves that refer to the cases with neutral gas occurs basically at the same energies obtained by using the Galactic diffusion coefficient (see { the} top panels in \autoref{fig:PhenW28}). At high energies, small differences with respect to the cases shown in \autoref{fig:PhenW28} are due to the different boundary condition{ s}: in the self-generation cases we assume free escape at some distance from the SNR, while in the phenomenological models the boundary escape is at infinity.

\section{Grammage}\label{sec:Grammage}

{ Although at present there is no compelling evidence of self-generation induced by CRs around SNRs, this phenomenon must occur at some level. The difficulty in observing, as discussed above, may be due to the complexity that is typical of the regions where these accelerators are located: such complexity would lead us to a multiplicity of phenomena taking place which would lead those regions to be excluded by an analysis such as the one presented here.}

{ The presence of regions of reduced diffusivity near sources leads to an increased residence time in these regions, and to the corresponding accumulation of grammage that is to be added to the grammage accumulated by CRs during their transport in the Galaxy.

This phenomenon is particularly important in the presence of dense MCs in the circum-source regions}, despite the fact that the time accumulated in such regions remains small compared to the escape time in the Galaxy \cite[]{2016PhRvD..94h3003D,2022A&A...660A..57R}. This is due to the larger { mean} density of { the} target gas that CRs traverse when MCs are present. The grammage accumulated in the absence of MCs was computed by \citet[]{2016PhRvD..94h3003D} and was found to be sizeable only for relatively large density of ionized gas { and absence of neutral hydrogen}, unlikely to be the case  in the disc of the Galaxy where SNRs are located. On the other hand, the confinement time is enhanced even if the ionized gas density is small. In this latter case, { as discussed below,} the accumulated grammage can be appreciable if MCs are present, even though surrounded by relatively rarefied ionized gas. 

Here, for simplicity, for the sole purpose of illustrating the effect, we assume that the CR protons are injected impulsively at $z=0$ at the edge of a { cylinder with a given radius, mimicing the transverse size of the source, depending on the gas density in the surrounding medium}.  

\begin{figure}
\includegraphics[scale=0.5]{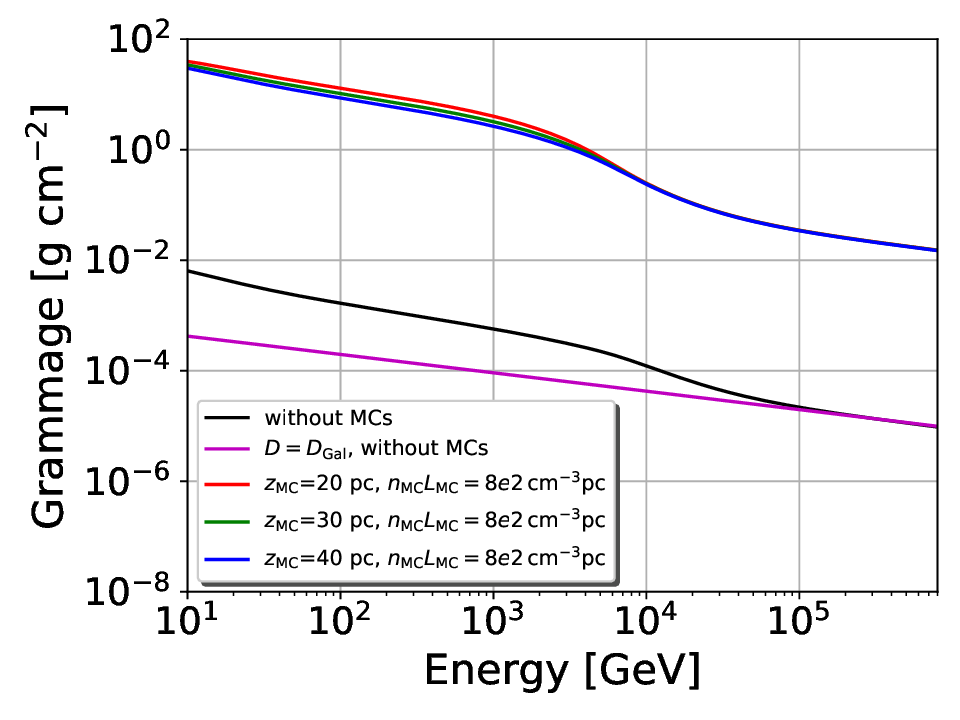}
\includegraphics[scale=0.5]{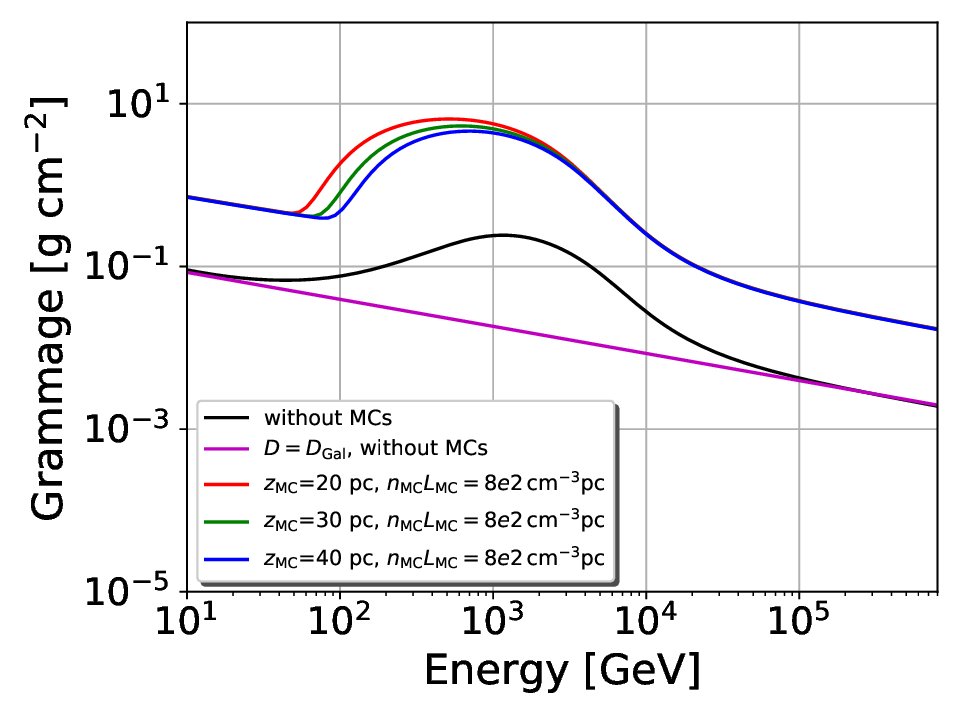}
\caption{CR grammage accumulated near a SNR for $n_{\rm i}=0.01$\,cm$^{-3}$ (left panel) and $n_{\rm i}$=$n_{\rm n}$=1\,cm$^{-3}$ (right panel). { The radius of the SNR is assumed to be 30 pc in the low density case (left panel) and 12 pc in the high density case (right panel)}. The MC is approximated to be a cylinder with radius 30 pc and centered at $z=z_{\rm MC}$. The free escape boundary is assumed to be $100$ pc away from the source.}
\label{fig:Gram}
\end{figure}

We solve the 1D diffusion-advection equation in the presence of self-generation, assuming a free escape boundary condition at a distance of $100$ pc from the SNR, and we compute the grammage as 
\begin{equation}
X_{\rm MC}(p)=\frac{\int_0^\infty \int_0^{L_{\rm c}} \rho_{\rm gas}(z) c f(p,z,t) {\rm d}z {\rm d}t}{\int_0^{L{\rm c}} f(p,z,0^{+}) {\rm d}z},
\end{equation}
where $\rho_{\rm gas}$ is the {\rm mass} density. The grammage accumulated by CRs in the near-SNR region is shown in \autoref{fig:Gram} for $n_i=0.01\rm cm^{-3}$ (left panel) and for $n_{\rm i}=n_{\rm n}=1\rm cm^{-3}$ (right panel). In both cases the cloud is assumed to be such that the MC density times the MC length $n_{MC}L_{MC}=800\rm cm^{-3}pc$ (clearly in a one dimensional problem this is the only relevant quantity). { The size of the source is set to 30 pc for the low density case (left panel) and 12 pc for the high density case (right panel).} The different curves refer to different values of $d_{\rm cl}$, the distance between the source and the center of the cloud. As expected, the latter parameter has little effect on the resulting grammage.  

In the left panel of \autoref{fig:Gram}, the grammage accumulated in the surrounding gas (without clouds) and using the Galactic diffusion coefficient is shown with a purple line. Even at $\sim 10$ GeV the corresponding grammage is negligibly small, $\sim 10^{-4}\rm g\, cm^{-2}$, as expected. When self-generation is taken into account, the residence time in the circum-source region increases and the corresponding grammage (black curve) increases correspondingly, but remains negligible. When a MC is present, on the other hand, the grammage becomes much larger, by roughly an amount $\sim n_{MC}L_{MC}/n_i L_c\sim 800$, where $L_{\rm c}$ is the location of the free escape boundary. { It is worth stressing that even in the absence of self-generation, in the presence of a MC the grammage can be boosted considerably, and a few $\rm g\,cm^{-2}$ can be accumulated at 10 GeV. Clearly, this does not mean that CRs from all SNRs undergo such phenomenon, but it is worth keeping in mind that the presence of MCs may affect the estimate of the grammage accumulated near sources.}

The situation when neutral gas is present (right panel in \autoref{fig:Gram}) is more interesting in that the effect of ion-neutral damping limits self-generation in an interesting manner as described below.

At very low energies, the effect of ion-neutral damping is so severe that the diffusion coefficient increases roughly to the Galactic value, hence in the absence of clouds the grammage overlaps with the purple line. When the cloud is present, the enhanced grammage is solely due to the fact that now the effective density in the region is $\sim n_{MC}L_{MC}/L_c\sim 8\rm cm^{-3}$, so that the grammage increases correspondingly. At very high energies the diffusion coefficient is again comparable with the one obtained with Galactic diffusion, due to the fact that the particles of such high energies escape too quickly for the effect of self-generation to be relevant. At energies between $\sim 100$ and $\sim 10^4$ GeV, on the other hand, ion-neutral damping is not sufficient to deplete self-generated waves completely, and hence the grammage shows a substantial enhancement, to $\sim 1-10~\rm g\,cm^{-2}$. Clearly, such an enhanced grammage would not be compatible with the observed grammage, as inferred from the B/C ratio and similar indicators, if this were a general phenomenon, around any source of CRs in the Galaxy. 

On the other hand, measurements of CR fluxes have reached such a high level of precision that it is conceivable that effects like that described above, even if subdominant, may be observable as deviations or anomalous behaviours of standard indicators of CR transport (for instance the B/C, B/O ratios). 

We mention in passing that there are alternative models of CR transport, mainly proposed to accommodate the positron excess in a purely secondary context, that advocate that most grammage might be accumulated in dense cocoons near sources \cite[]{Cowsik2014,Lipari2017}. { Although such models face daunting problems of self-consistency, especially when unstable secondary nuclei are included in the analysis, they forced us to consider the problem of CR transport near sources with more attention: given the accuracy of the data available at this time, even relatively small deviations from the standard picture of CR transport, such as those induced by CR transport around SNRs in the presence of MCs, might soon be unveiled.}

\section{Conclusion and discussion}\label{sec:Conclusions}

There are several reasons for exploring the possibility that CRs excite instabilities around sources, that in turn produce waves responsible for self-confinement. The first such reason is that there is now circumstantial evidence of reduced diffusion coefficients around some PWNe and { (less clearly)} SNRs, inferred from detailed $\gamma$-ray observations. The second reason is that this phenomenon, if sufficiently general, can impact on the grammage that CRs traverse while propagating in the Galaxy, a quantity that we infer from observing secondary to primary ratios in the cosmic radiation. For reasons explained above, SNRs and PWNe are expected to behave in qualitatively different ways and a dedicated investigation should be devoted to each of { the two} classes of sources. 

{ While this issue was investigated in previous articles \cite[]{2016PhRvD..94h3003D,2018MNRAS.474.1944D,2022A&A...660A..57R}, here we focused on several new aspects of the problem: 1) we identified a few instances of SNR-MC associations for which the $\gamma$-ray emission may flag the suppression of diffusivity and discussed a set of criteria to be adopted for this sake; 2) we analyzed the $\gamma$-ray emission from two SNRs, W28 and HB9, that are not expected to be particularly bright; 3) we investigated the effect of self-confinement on the grammage that CRs traverse while escaping the circum-source regions with special emphasis on the presence of MCs.} 

Extracting relevant physical information from the $\gamma$-ray emission associated with MCs around SNRs is particularly complex because of environmental issues: first, there are several cases in which the cloud is hit by the shock front of the SNR, a situation sometimes flagged by maser emission; in this case the acceleration process is severely affected and one should not necessarily expect that the $\gamma$-ray emission provides information about the diffusion coefficient in the medium surrounding the SNR. We avoid such situations. 

Second, in some cases, due to projection effects, it may be unclear whether a MC is associated with a given SNR. We try to avoid such situations as well, in that the results would be, to say the least, ambiguous.

In the attempt to identify some {\it clean cases}, in which the emission from (some of the) clouds is reasonably well associated with the production of pions by CRs diffusing away from a given SNR, { we chose one MC around SNR HB9 and two clouds around W28, as targets of our investigation and carried out the calculations in such cases}. 

A rough assessment of the possible need for a suppression in the diffusion coefficient around sources can be made by using a phenomenological calculation, in which the diffusion coefficient is parametrized as a fraction of a Galactic-like diffusion coefficient { \cite[]{2008A&A...481..401A,2013MNRAS.429.1643N,Bao2019,Bao2021}.} Meanwhile, a non-linear computation that includes self-generation and damping, clarifies whether the conditions are appropriate to expect a suppression of the diffusion coefficient as a result of the excitation of resonant streaming instability. 

We find that the case of MC R2 around the SNR HB9 is particularly useful: the diffusion coefficient required to account for the low energy part of the $\gamma$-ray emission is about two orders of magnitude smaller than the Galactic one and we show that it can be interpreted as a result of self-generation. { Unfortunately this apparently clear signature is made less evident by the degeneracy between this scenario and one based on a different picture of particle escape from the SNR: if the maximum energy for HB9 at the present age is as low as $\sim 100$ GeV, and the escape is assumed to be quasi-monochromatic, namely the particles escape the remnant from upstream only when their energy is within a narrow neighborhood of the maximum energy at that given time, then the effective spectrum of the particles in the near-source region has a low energy cut which reflects into a bending of the $\gamma$-ray spectrum, very similar to the observed one. We show that the $\gamma$-ray spectrum is equally well fit in the two scenarios, which leads to the impossibility to discriminate between them. We conclude that the evidence of suppression of diffusivity around HB9 is, at present, only circumstantial.}

The case of W28 is { even} more ambiguous: phenomenological models suggest that the evidence of small diffusion coefficient is limited to a 3D treatment of transport around this source, { confirming previous results by \cite{2013MNRAS.429.1643N}}. This is due to the fact that in three dimensions the density of CRs drops faster and for a given $\gamma$-ray emission this implies a smaller diffusion coefficient. On the other hand, a 3D treatment is justified only if the field lines in the region around the source are tangled on scales smaller than the source size. In this situation one can expect a smaller diffusion coefficient even in the absence of self-generation. 

In the one-dimensional case, our calculations of non-linear transport around W28 confirm that { one should not expect} appreciable self-generation, due to the presence of neutral gas in the region around W28. 

{ In conclusion, neither one of the SNR-MC associations considered here provides convincing proof that the diffusion coefficient is appreciably suppressed around these sources. One could argue that, besides being {\it cleaner} than other cases, neither SNR is particularly well suited for this purpose: in fact HB9 is known to be an underluminous SNR, so that the density of particles and their spatial gradients near the source could not be large. W28 is a relatively old SNR, hence possible non-linear transport would most likely be due to the particles injected in the past. Perhaps not surprisingly we cannot claim the identification of suppression of diffusivity in either case.}

{ Clearly, this does not imply that self-generation is not at work around SNRs, but rather that the most active regions that might be more suitable for this type of investigations are too complex and we have been unable, so far, to identify plausible candidates. If such regions do exist and the self-generation is at work, it is meaningful to ask what the implications would be in terms of grammage accumulated by CRs. While this problem was addressed in earlier studies \cite[]{2016PhRvD..94h3003D,2018MNRAS.474.1944D,2022A&A...660A..57R}, here we focused on the possible presence of MCs in the circum-source regions, possibly embedded in a medium that can be either dilute and ionized or denser and quasi-neutral. We find that the presence of MCs can substantially enhance the grammage accumulated near the source. In a realistic situation in which the MC is embedded in a partially ionized medium, the effect of ion-neutral damping is to reduce the level of self-generation for low energy particles, while self-generation remains effective for particles in the range $10^2-10^4$ GeV. Such a type of situation, if sufficiently common, would induce a feature in the same range of energy per nucleon in the secondary to primary ratios. Given the implications that this might have for the global CR transport, this phenomenon would definitely benefit from a dedicated investigation.}


\begin{acknowledgments}
{ We are very grateful to an anonymous referee for very useful comments concerning the scenario of escaping particles in a time dependent manner}. We also thank Qiang Yuan, Tian-yu Tu and Carmelo Evoli for helpful discussion and K.H.Yeung for providing $\gamma$-ray data. This work is supported by NSFC under grants 12173018, 12121003, and U1931204.
\end{acknowledgments}

\bibliography{sample631}{}

\begin{thebibliography}{}
\expandafter\ifx\csname natexlab\endcsname\relax\def\natexlab#1{#1}\fi
\providecommand{\url}[1]{\href{#1}{#1}}
\providecommand{\dodoi}[1]{doi:~\href{http://doi.org/#1}{\nolinkurl{#1}}}
\providecommand{\doeprint}[1]{\href{http://ascl.net/#1}{\nolinkurl{http://ascl.net/#1}}}
\providecommand{\doarXiv}[1]{\href{https://arxiv.org/abs/#1}{\nolinkurl{https://arxiv.org/abs/#1}}}

\bibitem[{{Abeysekara} {et~al.}(2017){Abeysekara}, {Albert}, {Alfaro},
  {Alvarez}, {{\'A}lvarez}, {Arceo}, {Arteaga-Vel{\'a}zquez}, {Avila Rojas},
  {Ayala Solares}, {Barber}, {Bautista-Elivar}, {Becerril}, {Belmont-Moreno},
  {BenZvi}, {Berley}, {Bernal}, {Braun}, {Brisbois}, {Caballero-Mora},
  {Capistr{\'a}n}, {Carrami{\~n}ana}, {Casanova}, {Castillo}, {Cotti},
  {Cotzomi}, {Couti{\~n}o de Le{\'o}n}, {De Le{\'o}n}, {De la Fuente},
  {Dingus}, {DuVernois}, {D{\'\i}az-V{\'e}lez}, {Ellsworth}, {Engel},
  {Enr{\'\i}quez-Rivera}, {Fiorino}, {Fraija}, {Garc{\'\i}a-Gonz{\'a}lez},
  {Garfias}, {Gerhardt}, {Gonz{\'a}lez Mu{\~n}oz}, {Gonz{\'a}lez}, {Goodman},
  {Hampel-Arias}, {Harding}, {Hern{\'a}ndez}, {Hern{\'a}ndez-Almada}, {Hinton},
  {Hona}, {Hui}, {H{\"u}ntemeyer}, {Iriarte}, {Jardin-Blicq}, {Joshi},
  {Kaufmann}, {Kieda}, {Lara}, {Lauer}, {Lee}, {Lennarz}, {Vargas},
  {Linnemann}, {Longinotti}, {Luis Raya}, {Luna-Garc{\'\i}a}, {L{\'o}pez-Coto},
  {Malone}, {Marinelli}, {Martinez}, {Martinez-Castellanos},
  {Mart{\'\i}nez-Castro}, {Mart{\'\i}nez-Huerta}, {Matthews},
  {Miranda-Romagnoli}, {Moreno}, {Mostaf{\'a}}, {Nellen}, {Newbold}, {Nisa},
  {Noriega-Papaqui}, {Pelayo}, {Pretz}, {P{\'e}rez-P{\'e}rez}, {Ren}, {Rho},
  {Rivi{\`e}re}, {Rosa-Gonz{\'a}lez}, {Rosenberg}, {Ruiz-Velasco}, {Salazar},
  {Salesa Greus}, {Sandoval}, {Schneider}, {Schoorlemmer}, {Sinnis}, {Smith},
  {Springer}, {Surajbali}, {Taboada}, {Tibolla}, {Tollefson}, {Torres},
  {Ukwatta}, {Vianello}, {Weisgarber}, {Westerhoff}, {Wisher}, {Wood},
  {Yapici}, {Yodh}, {Younk}, {Zepeda}, {Zhou}, {Guo}, {Hahn}, {Li}, \&
  {Zhang}}]{2017Sci...358..911A}
{Abeysekara}, A.~U., {Albert}, A., {Alfaro}, R., {et~al.} 2017, Science, 358,
  911, \dodoi{10.1126/science.aan4880}

\bibitem[{{Aguilar} {et~al.}(2016){Aguilar}, {Ali Cavasonza}, {Ambrosi},
  {Arruda}, {Attig}, {Aupetit}, {Azzarello}, {Bachlechner}, {Barao}, {Barrau},
  {Barrin}, {Bartoloni}, {Basara}, {Ba{\c{s}}e{\v{g}}mez-du Pree}, {Battarbee},
  {Battiston}, {Becker}, {Behlmann}, {Beischer}, {Berdugo}, {Bertucci},
  {Bindel}, {Bindi}, {Boella}, {de Boer}, {Bollweg}, {Bonnivard}, {Borgia},
  {Boschini}, {Bourquin}, {Bueno}, {Burger}, {Cadoux}, {Cai}, {Capell},
  {Caroff}, {Casaus}, {Castellini}, {Cervelli}, {Chae}, {Chang}, {Chen},
  {Chen}, {Chen}, {Cheng}, {Chou}, {Choumilov}, {Choutko}, {Chung}, {Clark},
  {Clavero}, {Coignet}, {Consolandi}, {Contin}, {Corti}, {Creus},
  {Crispoltoni}, {Cui}, {Dai}, {Delgado}, {Della Torre}, {Demakov},
  {Demirk{\"o}z}, {Derome}, {Di Falco}, {Dimiccoli}, {D{\'\i}az}, {von
  Doetinchem}, {Dong}, {Donnini}, {Duranti}, {D'Urso}, {Egorov}, {Eline},
  {Eronen}, {Feng}, {Fiandrini}, {Finch}, {Fisher}, {Formato}, {Galaktionov},
  {Gallucci}, {Garc{\'\i}a}, {Garc{\'\i}a-L{\'o}pez}, {Gargiulo}, {Gast},
  {Gebauer}, {Gervasi}, {Ghelfi}, {Giovacchini}, {Goglov}, {G{\'o}mez-Coral},
  {Gong}, {Goy}, {Grabski}, {Grandi}, {Graziani}, {Guo}, {Haino}, {Han}, {He},
  {Heil}, {Hoffman}, {Hsieh}, {Huang}, {Huang}, {Huh}, {Incagli}, {Ionica},
  {Jang}, {Jinchi}, {Kang}, {Kanishev}, {Kim}, {Kim}, {Kirn}, {Konak},
  {Kounina}, {Kounine}, {Koutsenko}, {Krafczyk}, {La Vacca}, {Laudi},
  {Laurenti}, {Lazzizzera}, {Lebedev}, {Lee}, {Lee}, {Leluc}, {Li}, {Li}, {Li},
  {Li}, {Li}, {Li}, {Li}, {Li}, {Li}, {Lim}, {Lin}, {Lipari}, {Lippert}, {Liu},
  {Liu}, {Lordello}, {Lu}, {Lu}, {Luebelsmeyer}, {Luo}, {Luo}, {Lv}, {Machate},
  {Majka}, {Ma{\~n}{\'a}}, {Mar{\'\i}n}, {Martin}, {Mart{\'\i}nez}, {Masi},
  {Maurin}, {Menchaca-Rocha}, {Meng}, {Mikuni}, {Mo}, {Morescalchi}, {Mott},
  {Nelson}, {Ni}, {Nikonov}, {Nozzoli}, {Oliva}, {Orcinha}, {Palmonari},
  {Palomares}, {Paniccia}, {Pauluzzi}, {Pensotti}, {Pereira}, {Picot-Clemente},
  {Pilo}, {Pizzolotto}, {Plyaskin}, {Pohl}, {Poireau}, {Putze}, {Quadrani},
  {Qi}, {Qin}, {Qu}, {R{\"a}ih{\"a}}, {Rancoita}, {Rapin}, {Ricol},
  {Rosier-Lees}, {Rozhkov}, {Rozza}, {Sagdeev}, {Sandweiss}, {Saouter},
  {Schael}, {Schmidt}, {Schulz von Dratzig}, {Schwering}, {Seo}, {Shan}, {Shi},
  {Siedenburg}, {Son}, {Song}, {Sun}, {Tacconi}, {Tang}, {Tang}, {Tao},
  {Tescaro}, {Ting}, {Ting}, {Tomassetti}, {Torsti}, {T{\"u}rko{\v{g}}lu},
  {Urban}, {Vagelli}, {Valente}, {Vannini}, {Valtonen}, {V{\'a}zquez Acosta},
  {Vecchi}, {Velasco}, {Vialle}, {Vitale}, {Vitillo}, {Wang}, {Wang}, {Wang},
  {Wang}, {Wang}, {Wang}, {Wei}, {Weng}, {Whitman}, {Wienkenh{\"o}ver}, {Wu},
  {Wu}, {Xia}, {Xiong}, {Xu}, {Yan}, {Yang}, {Yang}, {Yang}, {Yi}, {Yu}, {Yu},
  {Zeissler}, {Zhang}, {Zhang}, {Zhang}, {Zhang}, {Zhang}, {Zhang}, {Zheng},
  {Zhu}, {Zhuang}, {Zhukov}, {Zichichi}, {Zimmermann}, {Zuccon}, \& {AMS
  Collaboration}}]{2016PhRvL.117w1102A}
{Aguilar}, M., {Ali Cavasonza}, L., {Ambrosi}, G., {et~al.} 2016, \prl, 117,
  231102, \dodoi{10.1103/PhysRevLett.117.231102}

\bibitem[{{Aharonian} {et~al.}(2008){Aharonian}, {Akhperjanian}, {Bazer-Bachi},
  {Behera}, {Beilicke}, {Benbow}, {Berge}, {Bernl{\"o}hr}, {Boisson}, {Bolz},
  {Borrel}, {Braun}, {Brion}, {Brown}, {B{\"u}hler}, {Bulik}, {B{\"u}sching},
  {Boutelier}, {Carrigan}, {Chadwick}, {Chounet}, {Clapson}, {Coignet},
  {Cornils}, {Costamante}, {Degrange}, {Dickinson}, {Djannati-Ata{\"\i}},
  {Domainko}, {O'C. Drury}, {Dubus}, {Dyks}, {Egberts}, {Emmanoulopoulos},
  {Espigat}, {Farnier}, {Feinstein}, {Fiasson}, {F{\"o}rster}, {Fontaine},
  {Fukui}, {Funk}, {Funk}, {F{\"u}{\ss}ling}, {Gallant}, {Giebels},
  {Glicenstein}, {Gl{\"u}ck}, {Goret}, {Hadjichristidis}, {Hauser}, {Hauser},
  {Heinzelmann}, {Henri}, {Hermann}, {Hinton}, {Hoffmann}, {Hofmann},
  {Holleran}, {Hoppe}, {Horns}, {Jacholkowska}, {de Jager}, {Kendziorra},
  {Kerschhaggl}, {Kh{\'e}lifi}, {Komin}, {Kosack}, {Lamanna}, {Latham}, {Le
  Gallou}, {Lemi{\`e}re}, {Lemoine-Goumard}, {Lenain}, {Lohse}, {Martin},
  {Martineau-Huynh}, {Marcowith}, {Masterson}, {Maurin}, {McComb}, {Moderski},
  {Moriguchi}, {Moulin}, {de Naurois}, {Nedbal}, {Nolan}, {Olive}, {Orford},
  {Osborne}, {Ostrowski}, {Panter}, {Pedaletti}, {Pelletier}, {Petrucci},
  {Pita}, {P{\"u}hlhofer}, {Punch}, {Ranchon}, {Raubenheimer}, {Raue},
  {Rayner}, {Reimer}, {Renaud}, {Ripken}, {Rob}, {Rolland}, {Rosier-Lees},
  {Rowell}, {Rudak}, {Ruppel}, {Sahakian}, {Santangelo}, {Saug{\'e}},
  {Schlenker}, {Schlickeiser}, {Schr{\"o}der}, {Schwanke}, {Schwarzburg},
  {Schwemmer}, {Shalchi}, {Sol}, {Spangler}, {Stawarz}, {Steenkamp},
  {Stegmann}, {Superina}, {Takeuchi}, {Tam}, {Tavernet}, {Terrier}, {van
  Eldik}, {Vasileiadis}, {Venter}, {Vialle}, {Vincent}, {Vivier}, {V{\"o}lk},
  {Volpe}, {Wagner}, \& {Ward}}]{2008A&A...481..401A}
{Aharonian}, F., {Akhperjanian}, A.~G., {Bazer-Bachi}, A.~R., {et~al.} 2008,
  \aap, 481, 401, \dodoi{10.1051/0004-6361:20077765}

\bibitem[{{Amato} {et~al.}(2008){Amato}, {Blasi}, \&
  {Gabici}}]{2008MNRAS.385.1946A}
{Amato}, E., {Blasi}, P., \& {Gabici}, S. 2008, \mnras, 385, 1946,
  \dodoi{10.1111/j.1365-2966.2008.12876.x}

\bibitem[{{Bao} \& {Chen}(2021)}]{Bao2021}
{Bao}, Y., \& {Chen}, Y. 2021, \apj, 919, 32, \dodoi{10.3847/1538-4357/ac1581}

\bibitem[{{Bao} {et~al.}(2019){Bao}, {Liu}, \& {Chen}}]{Bao2019}
{Bao}, Y., {Liu}, S., \& {Chen}, Y. 2019, \apj, 877, 54,
  \dodoi{10.3847/1538-4357/ab1908}

\bibitem[{{Beck} {et~al.}(2016){Beck}, {Beck}, {Beck}, {Dolag}, {Strong}, \&
  {Nielaba}}]{2016JCAP...05..056B}
{Beck}, M.~C., {Beck}, A.~M., {Beck}, R., {et~al.} 2016, \jcap, 2016, 056,
  \dodoi{10.1088/1475-7516/2016/05/056}

\bibitem[{{Bell}(2004)}]{2004MNRAS.353..550B}
{Bell}, A.~R. 2004, \mnras, 353, 550, \dodoi{10.1111/j.1365-2966.2004.08097.x}

\bibitem[{{Bell} {et~al.}(2013){Bell}, {Schure}, {Reville}, \&
  {Giacinti}}]{2013MNRAS.431..415B}
{Bell}, A.~R., {Schure}, K.~M., {Reville}, B., \& {Giacinti}, G. 2013, \mnras,
  431, 415, \dodoi{10.1093/mnras/stt179}

\bibitem[{{Blasi}(2013)}]{2013A&ARv..21...70B}
{Blasi}, P. 2013, \aapr, 21, 70, \dodoi{10.1007/s00159-013-0070-7}

\bibitem[{{Blasi}(2019)}]{BlasiNCim}
---. 2019, Nuovo Cimento Rivista Serie, 42, 549,
  \dodoi{10.1393/ncr/i2019-10166-0}

\bibitem[{{Blasi} {et~al.}(2012){Blasi}, {Amato}, \& {Serpico}}]{Blasi2012}
{Blasi}, P., {Amato}, E., \& {Serpico}, P.~D. 2012, \prl, 109, 061101,
  \dodoi{10.1103/PhysRevLett.109.061101}

\bibitem[{{Brogan} {et~al.}(2006){Brogan}, {Gelfand}, {Gaensler}, {Kassim}, \&
  {Lazio}}]{2006ApJ...639L..25B}
{Brogan}, C.~L., {Gelfand}, J.~D., {Gaensler}, B.~M., {Kassim}, N.~E., \&
  {Lazio}, T.~J.~W. 2006, \apjl, 639, L25, \dodoi{10.1086/501500}

\bibitem[{{Caprioli} {et~al.}(2010){Caprioli}, {Amato}, \&
  {Blasi}}]{2010APh....33..160C}
{Caprioli}, D., {Amato}, E., \& {Blasi}, P. 2010, Astroparticle Physics, 33,
  160, \dodoi{10.1016/j.astropartphys.2010.01.002}

\bibitem[{{Cardillo} {et~al.}(2015){Cardillo}, {Amato}, \&
  {Blasi}}]{2015APh....69....1C}
{Cardillo}, M., {Amato}, E., \& {Blasi}, P. 2015, Astroparticle Physics, 69, 1,
  \dodoi{10.1016/j.astropartphys.2015.03.002}

\bibitem[{{Cioffi} {et~al.}(1988){Cioffi}, {McKee}, \&
  {Bertschinger}}]{1988ApJ...334..252C}
{Cioffi}, D.~F., {McKee}, C.~F., \& {Bertschinger}, E. 1988, \apj, 334, 252,
  \dodoi{10.1086/166834}

\bibitem[{{Cowsik} {et~al.}(2014){Cowsik}, {Burch}, \&
  {Madziwa-Nussinov}}]{Cowsik2014}
{Cowsik}, R., {Burch}, B., \& {Madziwa-Nussinov}, T. 2014, \apj, 786, 124,
  \dodoi{10.1088/0004-637X/786/2/124}

\bibitem[{{Cristofari} {et~al.}(2020){Cristofari}, {Blasi}, \&
  {Amato}}]{2020APh...12302492C}
{Cristofari}, P., {Blasi}, P., \& {Amato}, E. 2020, Astroparticle Physics, 123,
  102492, \dodoi{10.1016/j.astropartphys.2020.102492}

\bibitem[{{Cristofari} {et~al.}(2021){Cristofari}, {Blasi}, \&
  {Caprioli}}]{2021A&A...650A..62C}
{Cristofari}, P., {Blasi}, P., \& {Caprioli}, D. 2021, \aap, 650, A62,
  \dodoi{10.1051/0004-6361/202140448}

\bibitem[{{Cui} {et~al.}(2018){Cui}, {Yeung}, {Tam}, \&
  {P{\"u}hlhofer}}]{2018ApJ...860...69C}
{Cui}, Y., {Yeung}, P. K.~H., {Tam}, P.~H.~T., \& {P{\"u}hlhofer}, G. 2018,
  \apj, 860, 69, \dodoi{10.3847/1538-4357/aac37b}

\bibitem[{{D'Angelo} {et~al.}(2016){D'Angelo}, {Blasi}, \&
  {Amato}}]{2016PhRvD..94h3003D}
{D'Angelo}, M., {Blasi}, P., \& {Amato}, E. 2016, \prd, 94, 083003,
  \dodoi{10.1103/PhysRevD.94.083003}

\bibitem[{{D'Angelo} {et~al.}(2018){D'Angelo}, {Morlino}, {Amato}, \&
  {Blasi}}]{2018MNRAS.474.1944D}
{D'Angelo}, M., {Morlino}, G., {Amato}, E., \& {Blasi}, P. 2018, \mnras, 474,
  1944, \dodoi{10.1093/mnras/stx2828}

\bibitem[{{Evoli} {et~al.}(2018){Evoli}, {Linden}, \&
  {Morlino}}]{2018PhRvD..98f3017E}
{Evoli}, C., {Linden}, T., \& {Morlino}, G. 2018, \prd, 98, 063017,
  \dodoi{10.1103/PhysRevD.98.063017}

\bibitem[{{Evoli} {et~al.}(2020){Evoli}, {Morlino}, {Blasi}, \&
  {Aloisio}}]{Evoli2020}
{Evoli}, C., {Morlino}, G., {Blasi}, P., \& {Aloisio}, R. 2020, \prd, 101,
  023013, \dodoi{10.1103/PhysRevD.101.023013}

\bibitem[{{Farmer} \& {Goldreich}(2004)}]{2004ApJ...604..671F}
{Farmer}, A.~J., \& {Goldreich}, P. 2004, \apj, 604, 671,
  \dodoi{10.1086/382040}

\bibitem[{{Frail} {et~al.}(1996){Frail}, {Goss}, {Reynoso}, {Giacani}, {Green},
  \& {Otrupcek}}]{1996AJ....111.1651F}
{Frail}, D.~A., {Goss}, W.~M., {Reynoso}, E.~M., {et~al.} 1996, \aj, 111, 1651,
  \dodoi{10.1086/117904}

\bibitem[{{Gabici} {et~al.}(2007){Gabici}, {Aharonian}, \&
  {Blasi}}]{2007Ap&SS.309..365G}
{Gabici}, S., {Aharonian}, F.~A., \& {Blasi}, P. 2007, \apss, 309, 365,
  \dodoi{10.1007/s10509-007-9427-6}

\bibitem[{{Gabici} {et~al.}(2009){Gabici}, {Aharonian}, \&
  {Casanova}}]{2009MNRAS.396.1629G}
{Gabici}, S., {Aharonian}, F.~A., \& {Casanova}, S. 2009, \mnras, 396, 1629,
  \dodoi{10.1111/j.1365-2966.2009.14832.x}

\bibitem[{{Gabici} {et~al.}(2010){Gabici}, {Casanova}, {Aharonian}, \&
  {Rowell}}]{2010sf2a.conf..313G}
{Gabici}, S., {Casanova}, S., {Aharonian}, F.~A., \& {Rowell}, G. 2010, in
  SF2A-2010: Proceedings of the Annual meeting of the French Society of
  Astronomy and Astrophysics, ed. S.~{Boissier}, M.~{Heydari-Malayeri},
  R.~{Samadi}, \& D.~{Valls-Gabaud}, 313, \dodoi{10.48550/arXiv.1009.5291}

\bibitem[{{Giacinti} {et~al.}(2013){Giacinti}, {Kachelrie{\ss}}, \&
  {Semikoz}}]{2013PhRvD..88b3010G}
{Giacinti}, G., {Kachelrie{\ss}}, M., \& {Semikoz}, D.~V. 2013, \prd, 88,
  023010, \dodoi{10.1103/PhysRevD.88.023010}

\bibitem[{{Giacinti} \& {Kirk}(2018)}]{2018ApJ...863...18G}
{Giacinti}, G., \& {Kirk}, J.~G. 2018, \apj, 863, 18,
  \dodoi{10.3847/1538-4357/aacffb}

\bibitem[{{Hui} \& {Becker}(2007)}]{HuiBecker2007}
{Hui}, C.~Y., \& {Becker}, W. 2007, \aap, 467, 1209,
  \dodoi{10.1051/0004-6361:20066562}

\bibitem[{{Iqbal} {et~al.}(2009){Iqbal}, {Vahia}, {Masood}, \&
  {Ahmad}}]{2009JAHH...12...61I}
{Iqbal}, N., {Vahia}, M.~N., {Masood}, T., \& {Ahmad}, A. 2009, Journal of
  Astronomical History and Heritage, 12, 61

\bibitem[{{Joubert} {et~al.}(2016){Joubert}, {Castro}, {Slane}, \&
  {Gelfand}}]{2016ApJ...816...63J}
{Joubert}, T., {Castro}, D., {Slane}, P., \& {Gelfand}, J. 2016, \apj, 816, 63,
  \dodoi{10.3847/0004-637X/816/2/63}

\bibitem[{{Koldobskiy} {et~al.}(2021){Koldobskiy}, {Kachelrie{\ss}},
  {Lskavyan}, {Neronov}, {Ostapchenko}, \& {Semikoz}}]{2021PhRvD.104l3027K}
{Koldobskiy}, S., {Kachelrie{\ss}}, M., {Lskavyan}, A., {et~al.} 2021, \prd,
  104, 123027, \dodoi{10.1103/PhysRevD.104.123027}

\bibitem[{{Leahy} \& {Aschenbach}(1995)}]{1995A&A...293..853L}
{Leahy}, D.~A., \& {Aschenbach}, B. 1995, \aap, 293, 853

\bibitem[{{Leahy} \& {Tian}(2007)}]{2007A&A...461.1013L}
{Leahy}, D.~A., \& {Tian}, W.~W. 2007, \aap, 461, 1013,
  \dodoi{10.1051/0004-6361:20065895}

\bibitem[{{Leahy} \& {Williams}(2017)}]{2017AJ....153..239L}
{Leahy}, D.~A., \& {Williams}, J.~E. 2017, \aj, 153, 239,
  \dodoi{10.3847/1538-3881/aa6af6}

\bibitem[{{Li} \& {Chen}(2012)}]{2012MNRAS.421..935L}
{Li}, H., \& {Chen}, Y. 2012, \mnras, 421, 935,
  \dodoi{10.1111/j.1365-2966.2012.20270.x}

\bibitem[{{Lipari}(2017)}]{Lipari2017}
{Lipari}, P. 2017, \prd, 95, 063009, \dodoi{10.1103/PhysRevD.95.063009}

\bibitem[{{Lu} {et~al.}(2021){Lu}, {Guo}, {Kilian}, {Li}, {Huang}, \&
  {Liang}}]{2021ApJ...908..147L}
{Lu}, Y., {Guo}, F., {Kilian}, P., {et~al.} 2021, \apj, 908, 147,
  \dodoi{10.3847/1538-4357/abd406}

\bibitem[{{Lynds} \& {Oneil}(1985)}]{1985ApJ...294..578L}
{Lynds}, B.~T., \& {Oneil}, E.~J., J. 1985, \apj, 294, 578,
  \dodoi{10.1086/163325}

\bibitem[{{Malkov} {et~al.}(2013){Malkov}, {Diamond}, {Sagdeev}, {Aharonian},
  \& {Moskalenko}}]{Malkov2013}
{Malkov}, M.~A., {Diamond}, P.~H., {Sagdeev}, R.~Z., {Aharonian}, F.~A., \&
  {Moskalenko}, I.~V. 2013, \apj, 768, 73, \dodoi{10.1088/0004-637X/768/1/73}

\bibitem[{{Mukhopadhyay} \& {Linden}(2022)}]{2022PhRvD.105l3008M}
{Mukhopadhyay}, P., \& {Linden}, T. 2022, \prd, 105, 123008,
  \dodoi{10.1103/PhysRevD.105.123008}

\bibitem[{{Nava} \& {Gabici}(2013)}]{2013MNRAS.429.1643N}
{Nava}, L., \& {Gabici}, S. 2013, \mnras, 429, 1643,
  \dodoi{10.1093/mnras/sts450}

\bibitem[{{Nava} {et~al.}(2016){Nava}, {Gabici}, {Marcowith}, {Morlino}, \&
  {Ptuskin}}]{2016MNRAS.461.3552N}
{Nava}, L., {Gabici}, S., {Marcowith}, A., {Morlino}, G., \& {Ptuskin}, V.~S.
  2016, \mnras, 461, 3552, \dodoi{10.1093/mnras/stw1592}

\bibitem[{{Nava} {et~al.}(2019){Nava}, {Recchia}, {Gabici}, {Marcowith},
  {Brahimi}, \& {Ptuskin}}]{2019MNRAS.484.2684N}
{Nava}, L., {Recchia}, S., {Gabici}, S., {et~al.} 2019, \mnras, 484, 2684,
  \dodoi{10.1093/mnras/stz137}

\bibitem[{{Ohira} {et~al.}(2011){Ohira}, {Murase}, \&
  {Yamazaki}}]{2011MNRAS.410.1577O}
{Ohira}, Y., {Murase}, K., \& {Yamazaki}, R. 2011, \mnras, 410, 1577,
  \dodoi{10.1111/j.1365-2966.2010.17539.x}

\bibitem[{{Oka} \& {Ishizaki}(2022)}]{2022PASJ...74..625O}
{Oka}, T., \& {Ishizaki}, W. 2022, \pasj, 74, 625, \dodoi{10.1093/pasj/psac024}

\bibitem[{Olmi {et~al.}(2024)Olmi, Amato, Bandiera, \& Blasi}]{Olmi2024}
Olmi, B., Amato, E., Bandiera, R., \& Blasi, P. 2024, to appear in A\&A.
\newblock \doarXiv{2403.03616}

\bibitem[{{Olmi} \& {Bucciantini}(2019)}]{2019MNRAS.488.5690O}
{Olmi}, B., \& {Bucciantini}, N. 2019, \mnras, 488, 5690,
  \dodoi{10.1093/mnras/stz2089}

\bibitem[{{Ptuskin} {et~al.}(2009){Ptuskin}, {Strelnikova}, \&
  {Sveshnikova}}]{2009APh....31..284P}
{Ptuskin}, V.~S., {Strelnikova}, O.~N., \& {Sveshnikova}, L.~G. 2009,
  Astroparticle Physics, 31, 284, \dodoi{10.1016/j.astropartphys.2009.02.004}

\bibitem[{{Ptuskin} \& {Zirakashvili}(2003)}]{ZP2003}
{Ptuskin}, V.~S., \& {Zirakashvili}, V.~N. 2003, \aap, 403, 1,
  \dodoi{10.1051/0004-6361:20030323}

\bibitem[{{Ptuskin} {et~al.}(2008){Ptuskin}, {Zirakashvili}, \&
  {Plesser}}]{2008AdSpR..42..486P}
{Ptuskin}, V.~S., {Zirakashvili}, V.~N., \& {Plesser}, A.~A. 2008, Advances in
  Space Research, 42, 486, \dodoi{10.1016/j.asr.2007.12.007}

\bibitem[{{Reach} {et~al.}(2005){Reach}, {Rho}, \&
  {Jarrett}}]{2005ApJ...618..297R}
{Reach}, W.~T., {Rho}, J., \& {Jarrett}, T.~H. 2005, \apj, 618, 297,
  \dodoi{10.1086/425855}

\bibitem[{{Recchia} {et~al.}(2022){Recchia}, {Galli}, {Nava}, {Padovani},
  {Gabici}, {Marcowith}, {Ptuskin}, \& {Morlino}}]{2022A&A...660A..57R}
{Recchia}, S., {Galli}, D., {Nava}, L., {et~al.} 2022, \aap, 660, A57,
  \dodoi{10.1051/0004-6361/202142558}

\bibitem[{{Schroer} {et~al.}(2023){Schroer}, {Evoli}, \&
  {Blasi}}]{SchroerGeminga2023}
{Schroer}, B., {Evoli}, C., \& {Blasi}, P. 2023, arXiv e-prints,
  arXiv:2305.08019, \dodoi{10.48550/arXiv.2305.08019}

\bibitem[{{Schroer} {et~al.}(2021){Schroer}, {Pezzi}, {Caprioli}, {Haggerty},
  \& {Blasi}}]{2021ApJ...914L..13S}
{Schroer}, B., {Pezzi}, O., {Caprioli}, D., {Haggerty}, C., \& {Blasi}, P.
  2021, \apjl, 914, L13, \dodoi{10.3847/2041-8213/ac02cd}

\bibitem[{{Schroer} {et~al.}(2022){Schroer}, {Pezzi}, {Caprioli}, {Haggerty},
  \& {Blasi}}]{2022MNRAS.512..233S}
{Schroer}, B., {Pezzi}, O., {Caprioli}, D., {Haggerty}, C.~C., \& {Blasi}, P.
  2022, \mnras, 512, 233, \dodoi{10.1093/mnras/stac466}

\bibitem[{{Schure} \& {Bell}(2014)}]{2014MNRAS.437.2802S}
{Schure}, K.~M., \& {Bell}, A.~R. 2014, \mnras, 437, 2802,
  \dodoi{10.1093/mnras/stt2089}

\bibitem[{{Sezer} {et~al.}(2019){Sezer}, {Ergin}, {Yamazaki}, {Sano}, \&
  {Fukui}}]{2019MNRAS.489.4300S}
{Sezer}, A., {Ergin}, T., {Yamazaki}, R., {Sano}, H., \& {Fukui}, Y. 2019,
  \mnras, 489, 4300, \dodoi{10.1093/mnras/stz2461}

\bibitem[{{Skilling}(1971)}]{1971ApJ...170..265S}
{Skilling}, J. 1971, \apj, 170, 265, \dodoi{10.1086/151210}

\bibitem[{{Strong} {et~al.}(2007){Strong}, {Moskalenko}, \&
  {Ptuskin}}]{2007ARNPS..57..285S}
{Strong}, A.~W., {Moskalenko}, I.~V., \& {Ptuskin}, V.~S. 2007, Annual Review
  of Nuclear and Particle Science, 57, 285,
  \dodoi{10.1146/annurev.nucl.57.090506.123011}

\bibitem[{{Sudoh} {et~al.}(2019){Sudoh}, {Linden}, \&
  {Beacom}}]{2019PhRvD.100d3016S}
{Sudoh}, T., {Linden}, T., \& {Beacom}, J.~F. 2019, \prd, 100, 043016,
  \dodoi{10.1103/PhysRevD.100.043016}

\bibitem[{{Tothill} {et~al.}(2002){Tothill}, {White}, {Matthews}, {McCutcheon},
  {McCaughrean}, \& {Kenworthy}}]{2002ApJ...580..285T}
{Tothill}, N.~F.~H., {White}, G.~J., {Matthews}, H.~E., {et~al.} 2002, \apj,
  580, 285, \dodoi{10.1086/343068}

\bibitem[{{Truelove} \& {McKee}(1999)}]{1999ApJS..120..299T}
{Truelove}, J.~K., \& {McKee}, C.~F. 1999, \apjs, 120, 299,
  \dodoi{10.1086/313176}

\bibitem[{{Vel{\'a}zquez} {et~al.}(2002){Vel{\'a}zquez}, {Dubner}, {Goss}, \&
  {Green}}]{2002AJ....124.2145V}
{Vel{\'a}zquez}, P.~F., {Dubner}, G.~M., {Goss}, W.~M., \& {Green}, A.~J. 2002,
  \aj, 124, 2145, \dodoi{10.1086/342936}

\bibitem[{{White} \& {Long}(1991)}]{1991ApJ...373..543W}
{White}, R.~L., \& {Long}, K.~S. 1991, \apj, 373, 543, \dodoi{10.1086/170073}

\bibitem[{{Wiener} {et~al.}(2013){Wiener}, {Zweibel}, \&
  {Oh}}]{2013ApJ...767...87W}
{Wiener}, J., {Zweibel}, E.~G., \& {Oh}, S.~P. 2013, \apj, 767, 87,
  \dodoi{10.1088/0004-637X/767/1/87}

\bibitem[{{Yakovlev} \& {Pethick}(2004)}]{2004ARA&A..42..169Y}
{Yakovlev}, D.~G., \& {Pethick}, C.~J. 2004, \araa, 42, 169,
  \dodoi{10.1146/annurev.astro.42.053102.134013}

\bibitem[{{Zweibel} \& {Shull}(1982)}]{1982ApJ...259..859Z}
{Zweibel}, E.~G., \& {Shull}, J.~M. 1982, \apj, 259, 859,
  \dodoi{10.1086/160220}

\end{thebibliography}
\bibliographystyle{aasjournal}

\begin{center}
\begin{deluxetable}{cccc}
\tabletypesize{\footnotesize}
\tablecaption{Fitting Parameters for the linear phenomenological scenario \label{tab:Phen}}
\tablewidth{0pt}
\tablehead{
\\
Parameter     &    {HB9 R2} & {W28 MCA} & {W28 MCB}\\
}
\startdata
\hline
$t_{\rm age}$ (kyr)         & {7.0}   & \multicolumn{2}{c}{32}\\
$n_{\rm ISM}$ (cm$^{-3}$)   & {0.3}   & \multicolumn{2}{c}{5.0}\\
$M_{\rm ej}$ ($M_\odot$)    & {3.0}   & \multicolumn{2}{c}{6.0}\\
$t_{\rm ST}$ (kyr)       & {1.5}   & \multicolumn{2}{c}{0.6}\\
$t_{\rm PDS}$ (kyr)         & not entered & \multicolumn{2}{c}{5}\\
$\alpha_E$ 					& {2.4}   & \multicolumn{2}{c}{2.0}\\
$\rho_{\rm Inj}$ (pc, for 1D models) &{8.0}   & \multicolumn{2}{c}{7.0}\\
$E_{\rm SN}$ ($10^{51}$ erg)& {0.30}  & \multicolumn{2}{c}{1.0}\\
$d$ (kpc)                   & {0.60}  & \multicolumn{2}{c}{2.0}\\
\enddata
\tablenotetext{*}{Parameters used in phenomenological models.}
\end{deluxetable}
\end{center}

\begin{center}
\begin{deluxetable}{cccc}
\tabletypesize{\footnotesize}
\tablecaption{Fitting Parameters for the non-linear self-generated turbulence case\label{tab:par}}
\tablewidth{0pt}
\tablehead{
\\
Parameter     &   {HB9}   &   \multicolumn{2}{c}{W28}\\
}
\startdata
\multicolumn{4}{c}{SNR \& ISM} \\
\hline
$n_{\rm i}+n_{\rm n}$ (cm$^{-3}$) & {0.3} & \multicolumn{2}{c}{5.0}\\
$B_0$ ($\mu$G)				& {3.0}   & \multicolumn{2}{c}{3.0}\\
\hline
\multicolumn{4}{c}{MCs} \\
\hline
$z_1$ (pc)					& 10.0  & 10.0  & 10.0\\
$z_2$ (pc)					& 25.0  & 12.0  & 12.0\\
$m_{\rm MC}$ ($10^4M_\odot$)& (in legends)   & 4.3   & 5.7 \\
\enddata
\end{deluxetable}
\end{center}



\end{document}